\documentclass[]{aastex631}

\usepackage{multirow}
\usepackage{float}
\usepackage{booktabs}
\usepackage{txfonts}
\usepackage{rotating}
\begin{document}

\title{Characterizing Magnetic Properties of Young Protostars in Orion}

\author[0000-0001-7393-8583]{Bo Huang}
\email{huang@ice.csic.es}
\affiliation{Institut de Ciències de l'Espai (ICE-CSIC), Campus UAB, Can Magrans S/N, E-08193 Cerdanyola del Vallès, Catalonia, Spain}

\author[0000-0002-3829-5591]{Josep M. Girart}
\affiliation{Institut de Ciències de l'Espai (ICE-CSIC), Campus UAB, Can Magrans S/N, E-08193 Cerdanyola del Vallès, Catalonia, Spain}
\affiliation{Institut d’Estudis Espacials de Catalunya (IEEC), Campus del Baix Llobregat - UPC, Esteve Terradas 1, E-08860 Castelldefels, Catalonia, Spain}

\author[0000-0003-3017-4418]{Ian W. Stephens}
\affiliation{Department of Earth, Environment, and Physics, Worcester State University, Worcester, MA 01602, USA}

\author[0000-0002-2885-1806]{Philip C. Myers}
\affiliation{Center for Astrophysics $\mid$ Harvard $\&$ Smithsonian, 60 Garden Street, Cambridge, MA 02138, USA}

\author[0000-0003-2384-6589]{Qizhou Zhang}
\affiliation{Center for Astrophysics $\mid$ Harvard $\&$ Smithsonian, 60 Garden Street, Cambridge, MA 02138, USA}

\author[0000-0002-3583-780X]{Paulo Cortes}
\affiliation{National Radio Astronomy Observatory, 520 Edgemont Rd., Charlottesville, VA 22093, USA}
\affiliation{Joint ALMA Observatory, Alonso de Córdova 3107, Vitacura, Santiago, Chile}

\author[0000-0002-3078-9482]{\'Alvaro S\'anchez-Monge}
\affiliation{Institut de Ciències de l'Espai (ICE-CSIC), Campus UAB, Can Magrans S/N, E-08193 Cerdanyola del Vallès, Catalonia, Spain}
\affiliation{Institut d'Estudis Espacials de Catalunya (IEEC), c/Gran Capita, 2-4, E-08034 Barcelona, Catalonia, Spain}

\author[0000-0001-5811-0454]{Manuel Fern\'andez L\'opez}
\affiliation{Instituto Argentino de Radioastronomía (CCT-La Plata, CONICET; CICPBA), C.C. No. 5, 1894, Villa Elisa, Buenos Aires, Argentina}

\author[0000-0002-5714-799X]{Valentin J. M. Le Gouellec}
\affiliation{NASA Ames Research Center, Space Science and Astrobiology Division M.S. 245-6 Moffett Field, CA 94035, USA}
\affiliation{NASA Postdoctoral Program Fellow}

\author[0000-0001-7629-3573]{Tom Megeath}
\affiliation{Department of Astronomy, University of Toledo, Toledo, OH 43606, USA}

\author{Nadia M. Murillo}
\affiliation{Instituto de Astronom\'ia, Universidad Nacional Aut\'onoma de M\'exico, AP106, Ensenada CP 22830, B. C., M\'exico}
\affiliation{Star and Planet Formation Laboratory, RIKEN Cluster for Pioneering Research, Wako, Saitama 351-0198, Japan}

\author[0000-0003-2251-0602]{John M. Carpenter}
\affiliation{Joint ALMA Observatory, Av. Alonso de Córdova 3107, Vitacura, Santiago, Chile}

\author[0000-0002-7402-6487]{Zhi-Yun Li}
\affiliation{Astronomy Department, University of Virginia, Charlottesville, VA 22904, USA}

\author[0000-0002-4774-2998]{Junhao Liu}
\affiliation{National Astronomical Observatory of Japan, 2-21-1 Osawa, Mitaka, Tokyo 181-8588, Japan}

\author[0000-0002-4540-6587]{Leslie W. Looney}
\affiliation{Department of Astronomy, University of Illinois, 1002 West Green Street, Urbana, IL 61801, USA}

\author[0000-0001-7474-6874]{Sarah Sadavoy}
\affiliation{Department of Physics, Engineering and Astronomy, Queen’s University, 64 Bader Lane, Kingston, ON, K7L 3N6, Canada}

\author[0000-0003-3682-854X]{Nicole Karnath}
\affiliation{Space Science Institute, 4765 Walnut St, Suite B Boulder, CO 80301, USA}
\affiliation{Center for Astrophysics $\mid$ Harvard $\&$ Smithsonian, 60 Garden Street, Cambridge, MA 02138, USA}

\author[0000-0003-4022-4132]{Woojin Kwon}
\affiliation{Department of Earth Science Education, Seoul National University, 1 Gwanak-ro, Gwanak-gu, Seoul 08826, Republic of Korea}
\affiliation{SNU Astronomy Research Center, Seoul National University, 1 Gwanak-ro, Gwanak-gu, Seoul 08826, Republic of Korea}
\affiliation{The Center for Educational Research, Seoul National University, 1 Gwanak-ro, Gwanak-gu, Seoul 08826, Republic of Korea}

\begin{abstract}

The {\em B}-field Orion Protostellar Survey (BOPS) recently obtained polarimetric observations at 870 ${\rm \mu m}$ towards 61 protostars in the Orion molecular clouds with $\sim 1^{\prime\prime}$ spatial resolution using the Atacama Large Millimeter/submillimeter Array.
From the BOPS sample, we selected the 26 protostars with extended polarized emission within a radius of $\sim 6^{\prime\prime}$ (2400~au) around the protostar. This allows to have sufficient statistical polarization data to infer the magnetic field strength.
The magnetic field strength is derived using the Davis-Chandrasekhar-Fermi method.
The underlying magnetic field strengths are approximately 2.0~mG for protostars with a standard hourglass magnetic field morphology, which is higher than the values derived for protostars with rotated hourglass, spiral, and complex magnetic field configurations ($\lesssim1.0$~mG). This suggests that the magnetic field plays a more significant role in envelopes exhibiting a standard hourglass field morphology, and a value of $\gtrsim2.0$ mG would be required to maintain such a structure at these scales.
Furthermore, most protostars in the sample are slightly supercritical, with mass-to-flux ratios $\lesssim3.0$.
In particular, the mass-to-flux ratios for all protostars with a standard hourglass magnetic field morphology are lower than 3.0.
However, these ratios do not account for the contribution of the protostellar mass, which means they are likely significantly underestimated.
\end{abstract}


\section{Introduction} \label{sec:intro}

Magnetic fields (henceforth {\em B}-fields) are believed to play an important role during the star-forming process \citep[e.g.,][]{maury2022, pattle2023ppvii}.
Various theories on the formation and evolution of star clusters--ranging from those controlled by the {\em B}-field with ambipolar diffusion \citep[strong {\em B}-field model, e.g.,][]{inoue2013ins, van2014ism}, to those dominated by supersonic and super Alfv\'{e}nic turbulence \citep[weak {\em B}-field model, e.g.,][]{padoan2001ism, moechel2015ism}, or those driven by multiscale gravitational collapse \citep[global hierarchical collapse model, e.g.,][]{vazquez2019ghc,ramirez2022gravity}, have been developed in parallel during the last decades.
In the strong-field model, ambipolar diffusion enables the formation of ``supercritical'' dense cores, where the dominating force of gravity overcomes the magnetic support.
During the core collapse, field lines are dragged inward to form an hourglass morphology.
However, a recent numerical study by \cite{nacho2024magnetic} suggests that hourglass-shaped magnetic field structures can manifest in different magnetized environments. 

To gain a better understanding the role of {\em B}-fields, we can study them by observing polarized dust emission.
This is the most commonly used tracer of {\em B}-fields in star-forming regions, as ``radiative torques'' \citep[e.g.,][]{hoang2009grain, andersson2015interstellar} tend to align spinning and elongated dust grains with their long axes perpendicular to the ambient {\em B}-field direction.
Over the past decades, dust polarization observations carried out with (sub)millimeter interferometers have increasingly proven effective in mapping {\em B}-fields at the scales of cores ($\sim 10^{4}$ au) and envelopes ($\sim10^{2}$ to $10^{4}$ au) \citep[e.g.][]{girart1999detection, girart2013dr, zhang2014sma, cox2018alma, galametz2018sma, hull2019interferometric, le2020IMS, cortes2021magnetic, kwon2022obs}.
Additionally, hourglass-shaped {\em B}-fields have been observed on these scales \citep[e.g.,][]{girart2006magnetic, girart2009magnetic, stephens2013hourglass, qiu2014sma, le2019characterizing, hull2020understanding, huang2024magnetic}.

The Orion Molecular Cloud (OMC) star-forming complex is one of the most studied star-forming regions due to its proximity to Earth ($\sim$ 400 pc, \citealt{Kounkel2017dis}).
In a recent study, we used the Atacama Large Millimeter/submillimeter Array (ALMA) to observe the polarized dust emission and {\em B}-field structure in 61 protostars within the OMC, namely the {\em B}-field Orion Protostellar Survey \citep[BOPS, with a spatial resolution of $\sim1^{\prime\prime}$,][]{huang2024magnetic}.
The observations revealed not only the standard hourglass {\em B}-field structure (characterized by an inwardly constricted waist and outwardly flared lobes aligned parallel to the outflow axis, hereafter, std-hourglass), but also the rotated hourglass (morphologically identical to the std-hourglass but oriented perpendicularly to the outflow, hereafer, rot-hourglass), spiral patterns and complex {\em B}-field morphologies.
A follow-up study of BOPS \citep{huang2024b} found that protostars exhibiting std-hourglass {\em B}-field morphology tend to have smaller disk size and angle dispersion of the {\em B}-field compared to those with other {\em B}-field morphologies, particularly rot-hourglass morphologies, as well as spiral and complex configurations. 
This suggests that different {\em B}-field structures may indicate varying levels of significance of the field relative to turbulence and gravity.
Therefore, it is necessary to constrain the magnitude of the {\em B}-field, characterize the magnetic properties, and investigate the relationship between the {\em B}-field morphologies and the magnetization levels for the BOPS protostars.
Moreover, we found that 26 of the 61 BOPS protostellar envelopes have polarized dust emission detected with sufficient statistics to resolve the magnetic field structure, enabling a robust statistical examination and study of their magnetic properties.
Among them, six exhibit a std-hourglass {\em B}-field morphology, nine show a rot-hourglass structure, four display a spiral configuration, and the remaining seven are classified as complex.
In this paper, we aim to determine the influence of the {\em B}-field on these 26 protostellar envelopes and to gain a deeper understanding of the overall {\em B}-field properties on the envelope scales.
In Section \ref{sec:analysis}, we present the methods applied to estimate the {\em B}-field strength, followed by a detailed discussion in Section \ref{sec:discussions}.
Finally, we draw the main conclusions in Section \ref{sec:con}.

\section{Analysis} \label{sec:analysis}

The 870 $\mu$m (Band 7, \citealt{Mahieu2012alma}) dust polarization observations of BOPS were conducted with ALMA (2019.1.00086.S, PI: Ian Stephens), using its compact configurations C43-1 and C43-2.
The angular resolution is $\sim1^{\prime\prime}$, corresponding to a linear resolution of $\sim400$ au at a distance of $\sim400$ pc.
The largest angular scale (LAS) is approximately $8^{\prime\prime}$.
The C$^{17}$O (3-2) line was used to trace the velocity field on these scales.
For further details on the observation setup and data reduction, readers are referred to \cite{huang2024magnetic}.
The following sections present the estimation of physical envelope parameters and the total {\em B}-field strength for 26 protostars with extended polarized dust emission.
These results will then be used to examine the role of {\em B}-field on envelope scales.

\begin{table*}
\caption{Some parameters for the BOPS sample with extended polarized emission. Columns 2 to 8 present the envelope gas mass ($M_{\rm gas}$), hydrogen column density ($N_{\rm H_{2}}$), hydrogen number density ($n_{\rm H_{2}}$), volume density ($\rho$), non-thermal velocity dispersion ($\sigma_{\rm nth}$), {\em B}-field angle dispersion ($\delta \phi$), and the number of Nyquist-Sampled {\em B}-field segments ($N_{\rm NS}$), respectively. 
``Std-hourglass'', ``Rot-hourglass'', ``Spiral'', and ``Complex'' indicate protostars with different types of {\em B}-field structure of std-hourglass, rot-hourglass, spiral, and complex configurations, respectively.
$\sigma_{\rm nth}$ and $\delta\phi$ are taken from \cite{huang2024b}.}
\begin{center}
\label{Tab1:parameters}
\begin{tabular}{l c c c c c c c c c c c c c}
\hline
\hline
Name & $M_{\rm gas}$ (M$_{\odot}$) & ~ & $N_{\rm H_{2}}$ (cm$^{-2}$) & ~ & $n_{\rm H_{2}}$ (cm$^{-3}$) & ~ & $\rho$ (g cm$^{-3}$) & ~ & $\sigma_{\rm nth}$ (km~s$^{-1}$) & ~ & $\delta\phi$ ($^{\circ}$) & ~ & $N_{\rm NS}$  \\
\hline
\textbf{Std-hourglass} \\
HOPS-87 & 1.46 & ~~~~~ & 6.1E+23 & ~~~~~ & 2.6E+07 & ~~~~~ & 1.2E-16 & ~~~~~ & 0.46 $\pm$ 0.11 & ~~~~~ & 17.6 $\pm$ 1.4 & ~~~~~ & 86  \\
HOPS-359 & 0.62 & ~ & 2.6E+23 & ~ & 1.1E+07 & ~ & 5.1E-17 & ~ & 0.53 $\pm$ 0.07 & ~ & 29.2 $\pm$ 2.3 & ~ & 80 \\
HOPS-395 & 0.49 & ~ & 2.1E+23 & ~ & 8.6E+06 & ~ & 4.0E-17 & ~ & 0.31 $\pm$ 0.09 & ~ & 15.0 $\pm$ 2.0 & ~ & 29 \\
HOPS-400 & 1.01 & ~ & 4.3E+23 & ~ & 1.8E+07 & ~ & 8.3E-17 & ~ & 0.40 $\pm$ 0.12 & ~ & 18.2 $\pm$ 1.9 & ~ & 48  \\
HOPS-407 & 0.59 & ~ & 2.5E+23 & ~ & 1.0E+07 & ~ & 4.8E-17 & ~ & 0.43 $\pm$ 0.08 & ~ & 10.5 $\pm$ 1.0 & ~ & 53 \\
OMC1N-8-N & 0.31 & ~ & 1.3E+23 & ~ & 5.5E+06 & ~ & 2.6E-17 & ~ & 0.22 $\pm$ 0.01 & ~ & 20.6 $\pm$ 2.1 & ~ & 51 \\
\hline
\textbf{Rot-hourglass}\\
HH270IRS & 0.35 & ~ & 1.5E+23 & ~ & 6.1E+06 & ~ & 2.8E-17 & ~ & 0.42 $\pm$ 0.10 & ~ & 33.8 $\pm$ 4.1 & ~ & 35 \\
HOPS-78 & 0.48 & ~ & 2.0E+23 & ~ & 8.4E+06 & ~ & 3.9E-17 & ~ & 0.39 $\pm$ 0.10 & ~ & 40.2 $\pm$ 4.3 & ~ & 45 \\
HOPS-168 & 0.17 & ~ & 7.1E+22 & ~ & 3.0E+06 & ~ & 1.4E-17 & ~ & 0.51 $\pm$ 0.15 & ~ & 28.4 $\pm$ 3.4 & ~ & 28 \\
HOPS-169 & 0.47 & ~ & 2.0E+23 & ~ & 8.2E+06 & ~ & 3.9E-17 & ~ & 0.39 $\pm$ 0.09 & ~ & 32.9 $\pm$ 3.9 & ~ & 37 \\
HOPS-288 & 0.43 & ~ & 1.8E+23 & ~ & 7.6E+06 & ~ & 3.5E-17 & ~ & 0.61 $\pm$ 0.17 & ~ & 24.5 $\pm$ 2.9 & ~ & 37 \\
HOPS-317S & 1.95 & ~ & 8.2E+23 & ~ & 3.4E+07 & ~ & 1.6E-16 & ~ & 0.40 $\pm$ 0.10 & ~ & 33.7 $\pm$ 4.9 & ~ & 25 \\
HOPS-370 & 0.21 & ~ & 8.6E+22 & ~ & 3.6E+06 & ~ & 1.7E-17 & ~ & 0.68 $\pm$ 0.21 & ~ & 33.8 $\pm$ 3.1 & ~ & 62 \\
HOPS-409 & 0.23 & ~ & 9.8E+22 & ~ & 4.1E+06 & ~ & 1.9E-17 & ~ & 0.23 $\pm$ 0.03 & ~ & 17.0 $\pm$ 2.5 & ~ & 25 \\
OMC1N-4-5-ES & 0.28 & ~ & 1.2E+23 & ~ & 4.9E+06 & ~ & 2.3E-17 & ~ & 0.78 $\pm$ 0.27 & ~ & 34.6 $\pm$ 4.1 & ~ & 37 \\
\hline
\textbf{Spiral} \\
HOPS-182 & 0.42 & ~ & 1.8E+23 & ~ & 7.4E+06 & ~ & 3.4E-17 & ~ & 0.76 $\pm$ 0.37 & ~ & 46.4 $\pm$ 4.6 & ~ & 51 \\
HOPS-361N & 0.36 & ~ & 1.5E+23 & ~ & 6.3E+06 & ~ & 3.0E-17 & ~ & 0.77 $\pm$ 0.25 & ~ & 44.2 $\pm$ 3.4 & ~ & 86 \\
HOPS-361S & 0.23 & ~ & 9.8E+22 & ~ & 4.1E+06 & ~ & 1.9E-17 & ~ & 0.78 $\pm$ 0.33 & ~ & 43.1 $\pm$ 3.8 & ~ & 65 \\
HOPS-384 & 0.28 & ~ & 1.2E+23 & ~ & 4.9E+06 & ~ & 2.3E-17 & ~ & 0.74 $\pm$ 0.41 & ~ & 30.8 $\pm$ 2.2 & ~ & 98 \\
\hline
\textbf{Complex} \\
HOPS-12W & 0.30 & ~ & 1.2E+23 & ~ & 5.2E+06 & ~ & 2.4E-17 & ~ & 0.23 $\pm$ 0.03 & ~ & 39.7 $\pm$ 5.0 & ~ & 23 \\
HOPS-88 & 0.34 & ~ & 1.4E+23 & ~ & 6.0E+06 & ~ & 2.8E-17 & ~ & 0.45 $\pm$ 0.14 & ~ & 35.2 $\pm$ 4.3 & ~ & 35 \\
HOPS-373E & 0.43 & ~ & 1.8E+23 & ~ & 7.6E+06 & ~ & 3.6E-17 & ~ & 0.36 $\pm$ 0.08 & ~ & 28.9 $\pm$ 3.6 & ~ & 33 \\
HOPS-398 & 0.55 & ~ & 2.3E+23 & ~ & 9.7E+06 & ~ & 4.6E-17 & ~ & 0.38 $\pm$ 0.10 & ~ & 32.0 $\pm$ 5.0 & ~ & 20 \\
HOPS-399 & 1.28 & ~ & 5.4E+23 & ~ & 2.2E+07 & ~ & 1.0E-16 & ~ & 0.46 $\pm$ 0.09 & ~ & 46.4 $\pm$ 3.2 & ~ & 103 \\
OMC1N-4-5-EN & 0.18 & ~ & 7.6E+22 & ~ & 3.2E+06 & ~ & 1.5E-17 & ~ & 0.19 $\pm$ 0.01 & ~ & 16.4 $\pm$ 1.4 & ~ & 66 \\
OMC1N-6-7 & 0.38 & ~ & 1.6E+23 & ~ & 6.6E+06 & ~ & 3.1E-17 & ~ & 0.67 $\pm$ 0.18 & ~ & 39.2 $\pm$ 3.8 & ~ & 53 \\
\hline
\end{tabular}
\end{center}
\textbf{Note:} Typical uncertainties for the densities are: 40\% for $N_{\rm H_{2}}$, $n_{\rm H_{2}}$, and $\rho$ \citep{huang2024b}.
\end{table*}

\subsection{Physical Parameters}

The gas masses for these 26 protostars have been derived within the inner region of $R = 1200$ au \citep{huang2024b}, and will not be discussed further here to avoid redundancy.
This region covers an intensity detection of at least $\sim5\sigma$ in the Stokes {\em I} map.
The hydrogen column density $N_{\rm H_{2}}$ of the condensation is estimated using the following relation:
\begin{eqnarray}\label{eq1}
N_{\rm H_{2}} = \frac{M_{\rm gas}}{\pi R^{2}\mu_{\rm H_{2}}m_{\rm H}},
\end{eqnarray}
where $R=1200$ au is the adopted radius, $M_{\rm gas}$ is the envelope gas mass obtained from \cite{huang2024b}, $\mu_{\rm H_{2}} = 2.8$ is the mean molecular weight per hydrogen molecule \citep{kauffmann2008mambo}, and $m_{\rm H}$ is the mass of the hydrogen atom.
For a spherical region, the average hydrogen number density and volume density can be calculated as follows:
\begin{eqnarray}\label{eq2}
n_{\rm H_{2}} = \frac{3N_{\rm H_{2}}}{4R}, ~~~~ \rho = \mu_{\rm H_{2}}m_{\rm H}n_{\rm H_{2}}.
\end{eqnarray}
The physical parameters of $M_{\rm gas}$, $N_{\rm H_{2}}$, $n_{\rm H_{2}}$, and $\rho$ are listed in columns 2--5 of Table \ref{Tab1:parameters}.

\subsection{\textit{B}-field Strength}

The {\em B}-field in a star-forming region includes both the underlying and turbulent components of the {\em B}-field, which reflects the overall magnetic influence within the molecular cloud.
In star-forming regions, the turbulent component of the {\em B}-field strength contributes to local variability and instability within the cloud, while the underlying {\em B}-field is the ordered component of the {\em B}-field, which reflects the large-scale, coherent {\em B}-field that influences the initial stability of the cloud against gravitational collapse.
The typical approach to estimate the large-scale, ordered component of the {\em B}-field on the plane of the sky is the Davis-Chandrasekhar-Fermi \citep[DCF,][]{davis1951dcf, chandra1953magnetic} method, with its original, simplest form expressed as:
\begin{eqnarray}\label{eq3}
B_{\rm{u,~pos}}^{\rm DCF} \simeq \frac{\sqrt{4\pi\rho}}{\delta\phi} \sigma_{\rm nth}
\end{eqnarray}
in CGS units, where $\sigma_{\rm nth}$ is the non-thermal velocity dispersion, and $\delta\phi$ is the measured polarization angle dispersion, typically calculated as the standard deviation of the polarization position angles.
The field strength depends on the density, the non-thermal velocity dispersion, and the angle dispersion of polarization.
As emphasized by \cite{crutcher2004scuba}, it is important to note that these parameters should be estimated within the same scale, as the scale of the {\em B}-field angle dispersion must match that of the density estimation to improve the reliability of the DCF estimation.
The non-thermal velocity dispersion and the angle dispersion are obtained from \cite{huang2024b}, which were estimated within the same scale of inner 2400 au as that used for the density estimation.
Columns 6--7 list the values of the non-thermal velocity dispersion and the polarization position angle dispersion.

Notably, Equation \ref{eq3} is based on the assumptions that turbulent motions induce perturbations in a well-ordered mean {\em B}-field, and that the turbulent magnetic energy is small compared to the mean-field magnetic energy within the system, and the polarization position angle dispersions $\delta\phi$ should be lower than 25$^{\circ}$ to reliably reproduce the field strengths, as suggested by \cite{ostriker2001mag}.
They also proposed that the original DCF formula should be modified by multiplying a correction factor $Q_{\rm u,~pos}^{\rm DCF}\approx0.5$ to account for a more complex {\em B}-field morphology and density structure, which gives rise to more accurate measurements of the plane-of-sky {\em B}-field, expressed as:
\begin{eqnarray}\label{eq4}
B_{\rm u,~pos}^{\rm DCF} \simeq Q_{\rm u,~pos}^{\rm DCF}\frac{\sqrt{4\pi\rho}}{\delta\phi}\sigma_{\rm nth}.
\end{eqnarray}

For the cases of $\delta\phi \gtrsim 25^{\circ}$, the DCF method becomes less reliable.
Specifically, some studies discuss the limitations of the DCF method, particularly when dealing with large angle dispersions, the {\em B}-field strength is suggested to be significantly underestimated due to non-linear effects \citep[e.g.,][]{padoan1999cloud, ostriker2001mag, falceta2008DCF, crutcher2009magnetic, cho2016dcf, liu2021dcf, myers2024DCF}.
Therefore, several methods have been proposed to add corrections for the {\em B}-field angle dispersion to derive more accurate magnetic field strength.
For example, \cite{heitsch2001magnetic} (hereafter, Hei01) attempted to address the limitation of the small-angle approximation by replacing $\delta\phi$ by $\delta{\rm tan}(\phi-\overline{\phi})$ (here $\overline{\phi}$ is the mean value of the {\em B}-field position angle), and incorporating a geometric correction to avoid underestimation of the field in the super-Alfv\'{e}nic case.
Similarly, \cite{falceta2008DCF} (hereafter, Fal08) assumed that the field perturbation is a global property and substituted $\tan(\delta\phi$)$\sim\delta B/B_{\rm sky}$ for $\delta\phi$ in the denominator of Equation \ref{eq3}.
Additionally, \cite{skalidis2021dcf} suggested using $\sqrt{2\delta\phi}$, instead of $\delta\phi$, based on the assumption that the turbulent kinetic energy equals the magnetic energy fluctuations, without applying a correction factor.
However, several assumptions in \cite{skalidis2021dcf} are approximations, leading to large uncertainties in the estimation when no correction factor is applied \citep{liu2022dcf, myers2024DCF}.
Therefore, this method will not be discussed further in our study.
The two corrections of Hei01 and Fal08 can be expressed as:
\begin{eqnarray}\label{eq5}
B_{\rm u,~pos}^{\rm Hei01} = Q_{\rm u,~pos}^{\rm Hei01}\frac{\sqrt{4\pi\rho}}{\delta\tan(\phi-\overline{\phi})}\sigma_{\rm nth}(1+3\delta(\tan(\phi-\overline{\phi})^{2}))^{\frac{1}{4}},
\end{eqnarray}
\begin{eqnarray}\label{eq6}
B_{\rm u,~pos}^{\rm Fal08} = Q_{\rm u,~pos}^{\rm Fal08}\frac{\sqrt{4\pi\rho}\sigma_{\rm nth}}{\tan\delta\phi},
\end{eqnarray}
where $Q_{\rm u,~pos}^{\rm Hei01}$ and $Q_{\rm u,~pos}^{\rm Fal08}$ are the correction factors.
It is important to note that the correction factors in Hei01 and Fal08 are likely not constant and globally applicable in all environments.
Proper calibration is critical to increase the accuracy of the DCF method \citep[e.g.,][]{li2022magnetic}.
According to magnetohydrodynamics (MHD) turbulent simulations, the mean correction factor of Hei01 is $\sim0.3$ for fields stronger than the normalized field strength \citep{heitsch2001magnetic, liu2021dcf, chen2022dcf, myers2024DCF}.
For Fal08, multiplication by a factor 0.5--1 is recommended to account for projection from 3D to 2D angles, depending on the smoothness of angle variation along the line of sight \citep{falceta2008DCF, li2022magnetic, myers2024DCF}.
In this study, we set 0.5 for the correction factors of Hei01 and Fal08, which is within the empirical range, to implement the DCF method.
This choice is made because observations do not provide clear evidence for or against equipartition between turbulent kinetic and turbulent magnetic energy \citep[e.g.,][]{crutcher1999magnetic}.
Columns 5 of Table \ref{Tab2:strength} list the total field strengths of $B_{\rm u,~pos}^{\rm DCF}$ for 8 protostars with $\delta\phi\lesssim25^{\circ}$, while columns 6--7 list $B_{\rm u,~pos}^{\rm Hei01}$ and $B_{\rm u,~pos}^{\rm Fal08}$ for all protostars, respectively.
We note that the term $\delta{\rm tan}(\phi-\overline{\phi})$ in Euqation~\ref{eq5} has no physical meaning when $\phi-\overline{\phi}$ is close to 90$^{\circ}$. This produces unreasonably low values, which is clearly the case of HOPS-361N.

\begin{table*}
\caption{Parameters of total {\em B}-field strength. Column 2 gives the {\em B}-field angle dispersion of $\delta \phi$, which is used in Equation \ref{eq4}; columns 3--4 list the corrected {\em B}-field angle dispersion, which are used in Equation \ref{eq5} and Equation \ref{eq6}, respectively.
Columns 5--7 give the total {\em B}-field strengths using the standard DCF correction given in Equation \ref{eq4} \citep{ostriker2001mag}, and the corrected angle dispersion of Equation \ref{eq5} \citep{heitsch2001magnetic} and Equation \ref{eq6} \citep{falceta2008DCF}, respectively.
Columns 8--10 gives the mass-to-flux ratio using the total {\em B}-field strength listed in columns 5--7.
``Std-hourglass'', ``Rot-hourglass'', ``Spiral'', and ``Complex'' indicate protostars with different types of {\em B}-field structure of std-hourglass, rot-hourglass, spiral, and complex configurations, respectively.
The missing values in columns 5 and 8 because $\delta\phi\gtrsim25^{\circ}$, leading to less reliable values when using the standard DCF method.}
\begin{center}
\label{Tab2:strength}
\begin{tabular}{l c c c c c c c c c}
\hline
\hline
Name & $\delta\phi$ & $\frac{\delta\tan(\phi-\overline{\phi})}{(1+3\delta(\tan(\phi-\overline{\phi})^{2}))^{0.25}}$ & $\tan(\delta\phi)$ & $B_{\rm u,~pos}^{\rm DCF}$ & $B_{\rm u,~pos}^{\rm Hei01}$ & $B_{\rm u,~pos}^{\rm Fal08}$ & $\lambda_{\rm u,~pos}^{\rm DCF}$ & $\lambda_{\rm u,~pos}^{\rm Hei01}$ & $\lambda_{\rm u,~pos}^{\rm Fal08}$ \\
~ & (rad) & (rad) & (rad) & (mG) & (mG) & (mG) & ~ & ~  \\
\hline
\textbf{Std-hourglass} \\
HOPS-87 & 0.31 $\pm$ 0.02 & 0.31 $\pm$ 0.04 & 0.32 $\pm$ 0.03 & 2.9 $\pm$ 0.9 & 2.9 $\pm$ 1.0 & 2.8 $\pm$ 0.9 & 2.0 $\pm$ 0.6 & 2.0 $\pm$ 0.7 & 2.0 $\pm$ 0.7 \\
HOPS-359 & 0.51 $\pm$ 0.04 & 1.32 $\pm$ 6.43 & 0.56 $\pm$ 0.05 & / & 0.5 $\pm$ 2.5 & 1.0 $\pm$ 0.3 & / & 4.8 $\pm$ 23.3 & 2.5 $\pm$ 0.8 \\
HOPS-395 & 0.26 $\pm$ 0.03 & 0.26 $\pm$ 0.05 & 0.27 $\pm$ 0.04 & 1.3 $\pm$ 0.5 & 1.3 $\pm$ 0.3 & 1.3 $\pm$ 0.5 & 1.5 $\pm$ 0.5 & 1.5 $\pm$ 0.6 & 1.5 $\pm$ 0.6 \\
HOPS-400 & 0.32 $\pm$ 0.03 & 0.32 $\pm$ 0.06 & 0.33 $\pm$ 0.04 & 2.1 $\pm$ 0.8 & 2.0 $\pm$ 0.8 & 2.0 $\pm$ 0.8 & 1.9 $\pm$ 0.7 & 1.9 $\pm$ 0.8 & 2.0 $\pm$ 0.8 \\
HOPS-407 & 0.18 $\pm$ 0.03 & 0.18 $\pm$ 0.03 & 0.19 $\pm$ 0.02 & 2.9 $\pm$ 0.9 & 2.9 $\pm$ 0.9 & 2.9 $\pm$ 0.8 & 0.8 $\pm$ 0.2 & 0.8 $\pm$ 0.2 & 0.8 $\pm$ 0.2 \\
OMC1N-8-N & 0.36 $\pm$ 0.04 & 0.37 $\pm$ 0.09 & 0.38 $\pm$ 0.04 & 0.6 $\pm$ 0.1 & 0.5 $\pm$ 0.2 & 0.5 $\pm$ 0.1 & 2.2 $\pm$ 0.5 & 2.3 $\pm$ 0.7 & 2.3 $\pm$ 0.5 \\
\hline
\textbf{Rot-hourglass}\\
HH270IRS & 0.59 $\pm$ 0.07 & 0.59 $\pm$ 0.17 & 0.67 $\pm$ 0.10 & / & 0.7 $\pm$ 0.3 & 0.6 $\pm$ 0.2 & / & 2.0 $\pm$ 0.9 & 2.3 $\pm$ 0.8 \\
HOPS-78 & 0.70 $\pm$ 0.08 & 0.68 $\pm$ 0.22 & 0.85 $\pm$ 0.13 & / & 0.6 $\pm$ 0.3 & 0.5 $\pm$ 0.2 & / & 3.0 $\pm$ 1.4 & 3.7 $\pm$ 1.3 \\
HOPS-168 & 0.50 $\pm$ 0.07 & 0.48 $\pm$ 0.19 & 0.54 $\pm$ 0.09 & / & 0.7 $\pm$ 0.4 & 0.6 $\pm$ 0.2 & / & 0.9 $\pm$ 0.5 & 1.1 $\pm$ 0.4 \\
HOPS-169 & 0.57 $\pm$ 0.07 & 1.02 $\pm$ 2.42 & 0.65 $\pm$ 0.10 & / & 0.4 $\pm$ 1.0 & 0.7 $\pm$ 0.2 & / & 4.4 $\pm$ 10.5 & 2.8 $\pm$ 0.9 \\
HOPS-288 & 0.43 $\pm$ 0.05 & 0.45 $\pm$ 0.17 & 0.46 $\pm$ 0.06 & 1.5 $\pm$ 0.6 & 1.4 $\pm$ 0.7 & 1.4 $\pm$ 0.5 & 1.1 $\pm$ 0.4 & 1.2 $\pm$ 0.6 & 1.2 $\pm$ 0.5 \\
HOPS-317S & 0.59 $\pm$ 0.08 & 0.62 $\pm$ 0.14 & 0.67 $\pm$ 0.12 & / & 1.4 $\pm$ 0.5 & 1.3 $\pm$ 0.5 & / & 5.4 $\pm$ 2.1 & 5.7 $\pm$ 2.1 \\
HOPS-370 & 0.59 $\pm$ 0.05 & 0.58 $\pm$ 0.19 & 0.67 $\pm$ 0.08 & / & 0.8 $\pm$ 0.4 & 0.7 $\pm$ 0.3 & / & 1.0 $\pm$ 0.5 & 1.1 $\pm$ 0.4 \\
HOPS-409 & 0.30 $\pm$ 0.04 & 0.31 $\pm$ 0.09 & 0.31 $\pm$ 0.06 & 0.6 $\pm$ 0.2 & 0.6 $\pm$ 0.2 & 0.6 $\pm$ 0.2 & 1.5 $\pm$ 0.4 & 1.6 $\pm$ 0.6 & 1.6 $\pm$ 0.5 \\
OMC1N-4-5-ES & 0.60 $\pm$ 0.07 & 0.56 $\pm$ 0.20 & 0.69 $\pm$ 0.11 & / & 1.2 $\pm$ 0.6 & 1.0 $\pm$ 0.4 & / & 0.9 $\pm$ 0.5 & 1.1 $\pm$ 0.5 \\
\hline
\textbf{Spiral}\\
HOPS-182 & 0.81 $\pm$ 0.08 & 1.12 $\pm$ 3.73 & 1.05 $\pm$ 0.17 & / & 0.7 $\pm$ 2.4 & 0.8 $\pm$ 0.4 & / & 2.3 $\pm$ 7.8 & 2.2 $\pm$ 1.2 \\
HOPS-361N & 0.77 $\pm$ 0.06 & 17 $\pm$ 14859 & 0.97 $\pm$ 0.12 & / & 0.04 $\pm$ 38.0 & 0.8 $\pm$ 0.3 & / & 32 $\pm$ 28365 & 1.9 $\pm$ 0.7 \\
HOPS-361S & 0.75 $\pm$ 0.07 & 2.13 $\pm$ 20.73 & 0.94 $\pm$ 0.12 & / & 0.3 $\pm$ 2.8 & 0.6 $\pm$ 0.3 & / & 3.2 $\pm$ 31.2 & 1.4 $\pm$ 0.7 \\
HOPS-384 & 0.54 $\pm$ 0.04 & 0.53 $\pm$ 0.10 & 0.60 $\pm$ 0.05 & / & 1.2 $\pm$ 0.7 & 1.0 $\pm$ 0.6 & / & 0.9 $\pm$ 0.6 & 1.0 $\pm$ 0.6 \\
\hline
\textbf{Complex}\\
HOPS-12W & 0.69 $\pm$ 0.09 & 1.46 $\pm$ 9.41 & 0.83 $\pm$ 0.15 & / & 0.1 $\pm$ 0.9 & 0.2 $\pm$ 0.1 & / & 8.2 $\pm$ 53.2 & 4.7 $\pm$ 1.4 \\
HOPS-88 & 0.61 $\pm$ 0.08 & 0.58 $\pm$ 0.27 & 0.71 $\pm$ 0.11 & / & 0.7 $\pm$ 0.4 & 0.6 $\pm$ 0.2 & / & 1.8 $\pm$ 1.1 & 2.2 $\pm$ 0.9 \\
HOPS-373E & 0.50 $\pm$ 0.06 & 0.50 $\pm$ 0.13 & 0.55 $\pm$ 0.08 & / & 0.8 $\pm$ 0.3 & 0.7 $\pm$ 0.2 & / & 2.2 $\pm$ 0.9 & 2.4 $\pm$ 0.8 \\
HOPS-398 & 0.56 $\pm$ 0.09 & 0.91 $\pm$ 2.16 & 0.62 $\pm$ 0.12 & / & 0.5 $\pm$ 1.2 & 0.7 $\pm$ 0.3 & / & 4.3 $\pm$ 10.4 & 3.0 $\pm$ 1.1 \\
HOPS-399 & 0.81 $\pm$ 0.06 & 0.79 $\pm$ 0.59 & 1.05 $\pm$ 0.12 & / & 1.1 $\pm$ 0.8 & 0.8 $\pm$ 0.2 & / & 4.7 $\pm$ 3.7 & 6.2 $\pm$ 1.9 \\
OMC1N-4-5-EN & 0.29 $\pm$ 0.02 & 0.34 $\pm$ 0.11 & 0.29 $\pm$ 0.03 & 0.5 $\pm$ 0.1 & 0.4 $\pm$ 0.2 & 0.4 $\pm$ 0.1 & 1.5 $\pm$ 0.3 & 1.8 $\pm$ 0.7 & 1.6 $\pm$ 0.4 \\
OMC1N-6-7 & 0.68 $\pm$ 0.07 & 1.73 $\pm$ 12.03 & 0.82 $\pm$ 0.11 & / & 0.4 $\pm$ 2.7 & 0.8 $\pm$ 0.3 & / & 3.9 $\pm$ 26.8 & 1.8 $\pm$ 0.7 \\
\hline
\end{tabular}
\end{center}
\end{table*}

\begin{figure*}
\centering
\includegraphics[clip=true,trim=0.3cm 0.5cm 0.3cm 0.3cm,width=0.48 \textwidth]{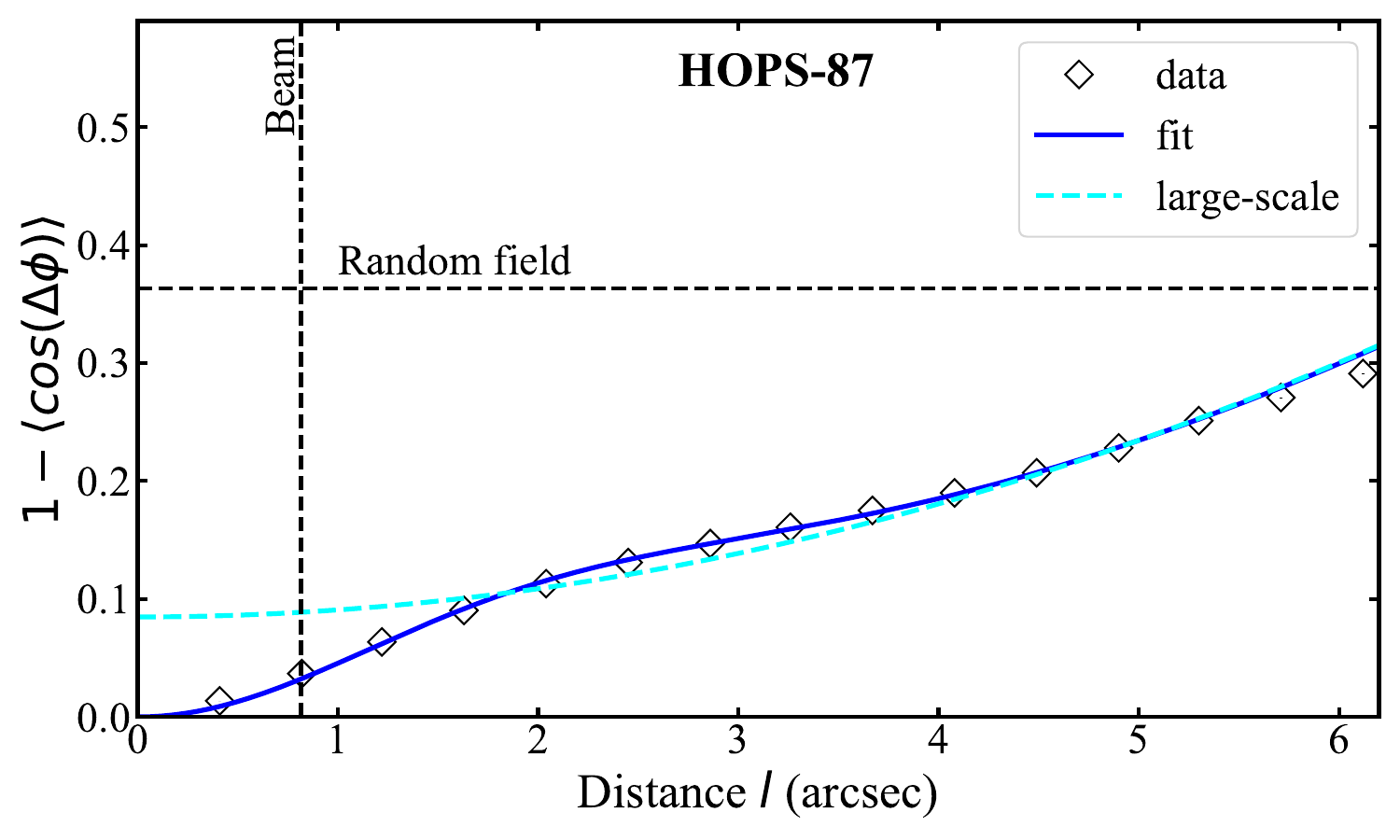}~~~~
\includegraphics[clip=true,trim=0.3cm 0.5cm 0.3cm 0.3cm,width=0.48 \textwidth]{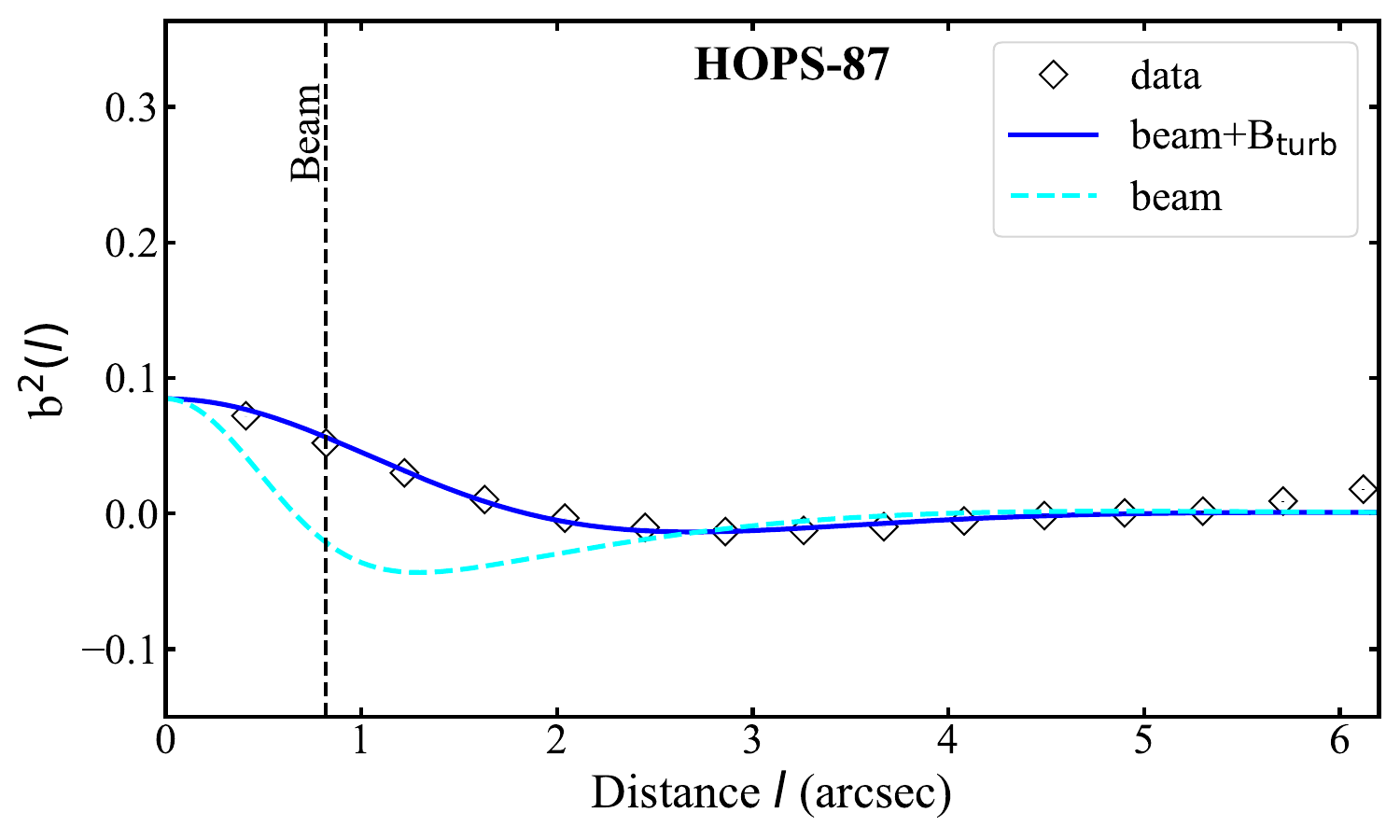}
~\\
~\\
\includegraphics[clip=true,trim=0.3cm 0.5cm 0.3cm 0.3cm,width=0.48 \textwidth]{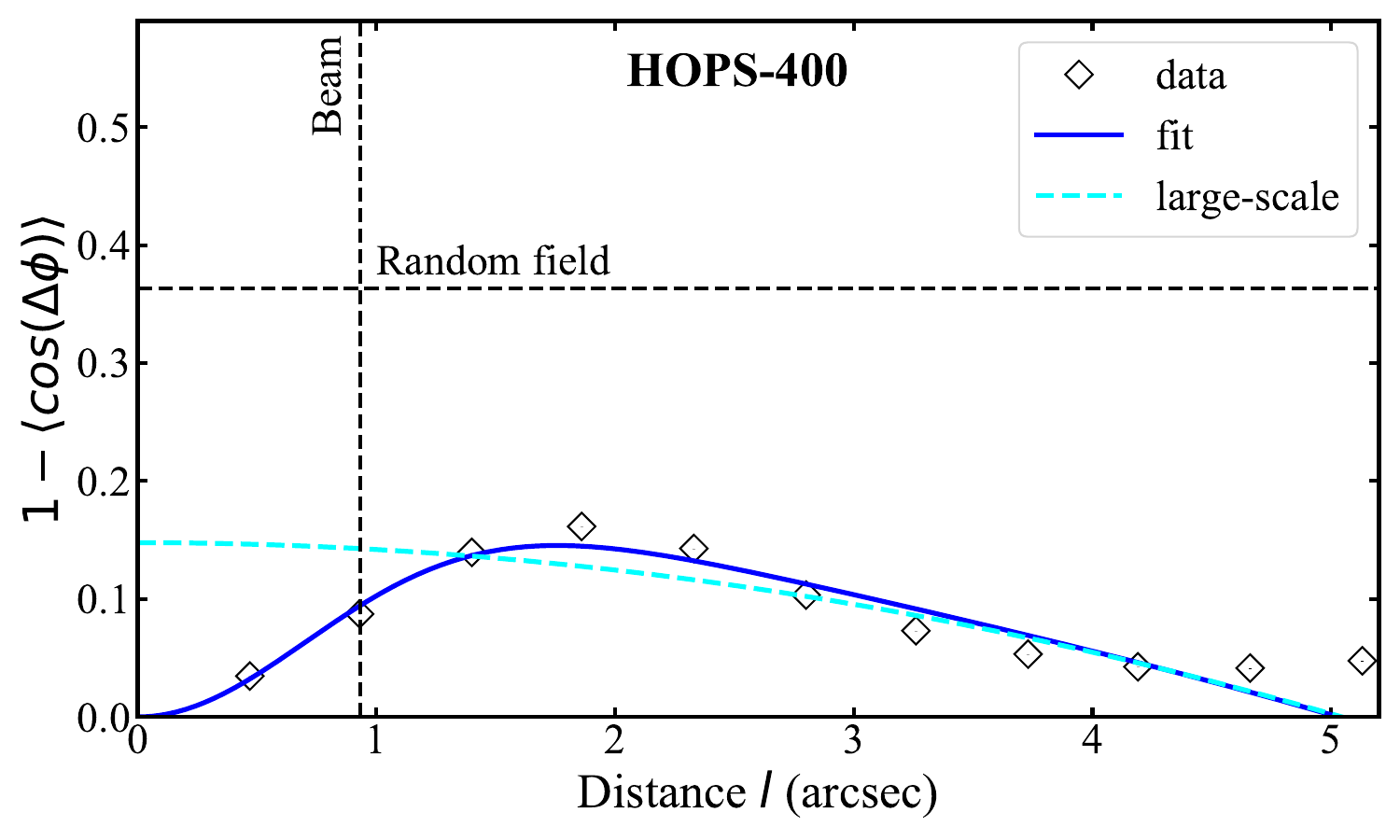} ~~~
\includegraphics[clip=true,trim=0.3cm 0.5cm 0.3cm 0.3cm,width=0.48 \textwidth]{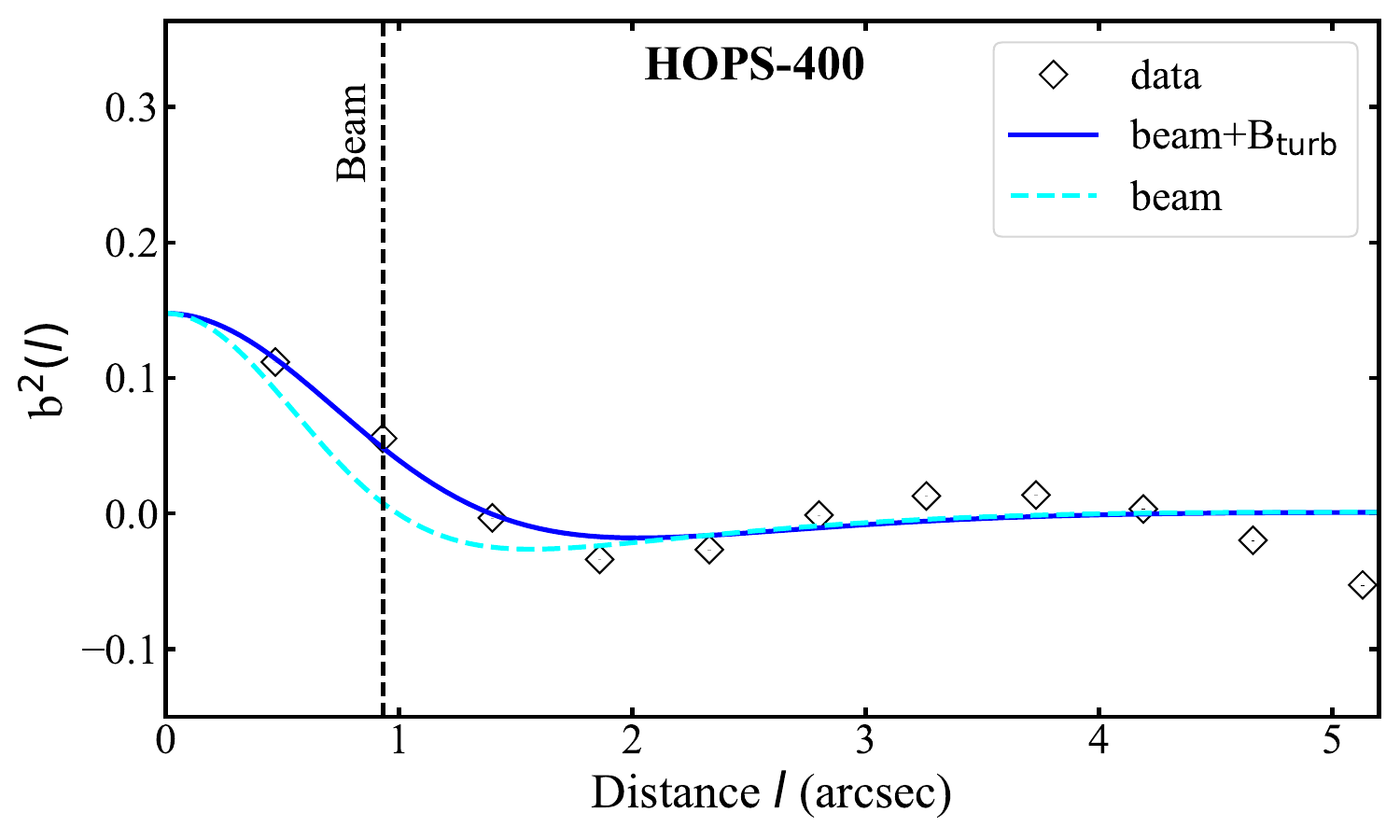}
~\\
~\\
\includegraphics[clip=true,trim=0.3cm 0.5cm 0.3cm 0.3cm,width=0.48 \textwidth]{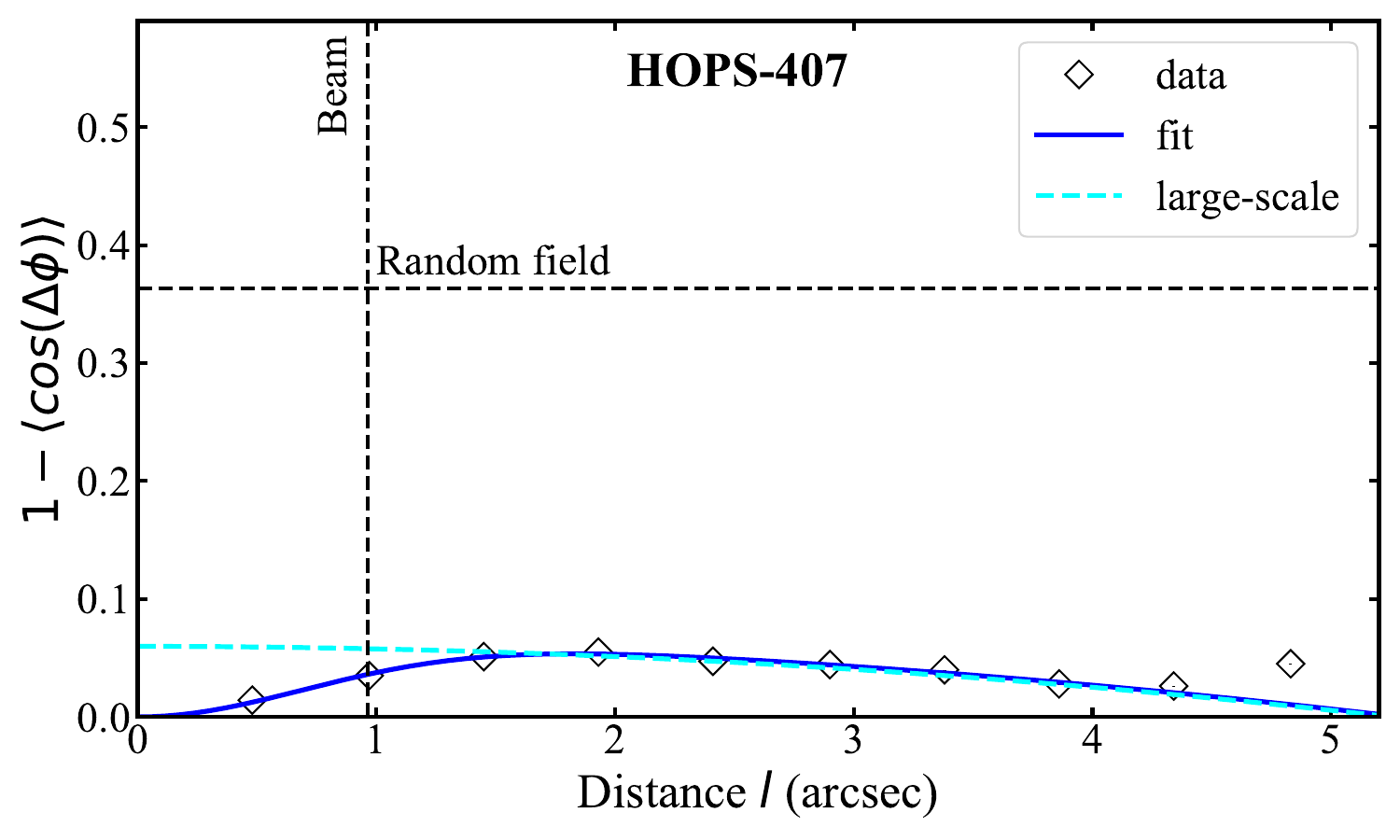}~~~~
\includegraphics[clip=true,trim=0.3cm 0.5cm 0.3cm 0.3cm,width=0.48 \textwidth]{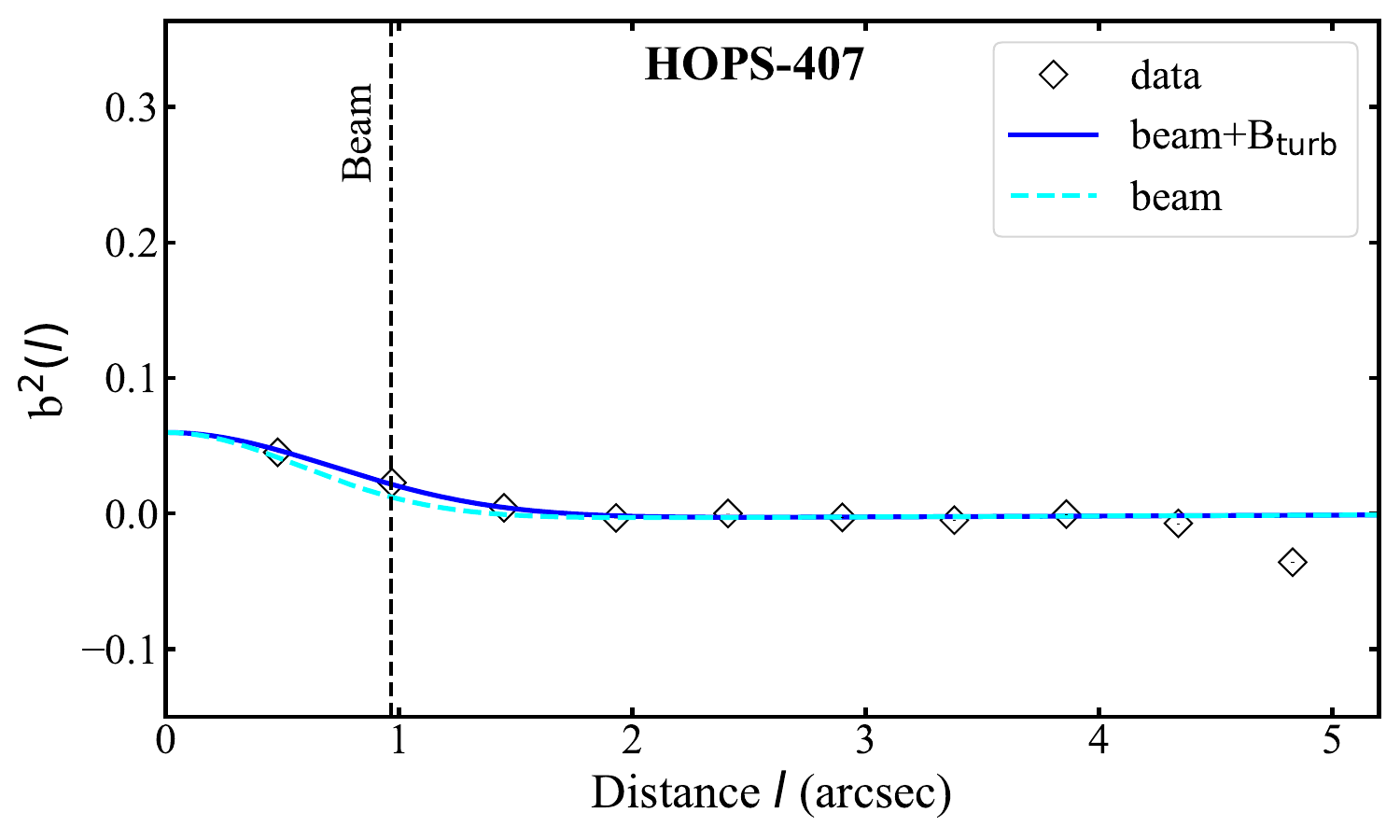}
\caption{Left panels: Angular dispersion function.
Angle dispersion segments are shown as diamond symbols.
The blue line shows the fitted ADF, while the cyan dashed line shows the large-scale component $(1- \langle {\rm cos}[\Delta\Phi(l)]\rangle -b^{2}(l))$ of the best fit. 
Horizontal and vertical dashed lines indicate the value corresponding to a random field and the synthesized beam size, respectively.
Right panels: Correlated component $(b^{2}(l))$ of the ADF. 
The correlated components of the best fit are shown in the blue lines.
The correlated components solely due to the beam are shown in the cyan dashed lines.}
\label{fig:adf}
\end{figure*}

An alternative approach has been proposed to modify the standard DCF method, aiming to more accurately quantify the angle dispersion inherent to the DCF formula. 
This is achieved through the angular dispersion function (ADF) analysis \citep{hildebrand2009dispersion, houde2009dispersion, houde2016dispersion}.
Specifically, \cite{houde2016dispersion} derived the ADF for polarimetric images obtained from an interferometer, accounting for variations in the large-scale {\em B}-field, the effects of signal integration along the line of sight and within the beam, and the large-scale filtering effect of interferometers. 
The ADF analysis, as described by \cite{houde2016dispersion}, is based on the following equation:
\begin{eqnarray}\label{eq7}
1-\langle \cos[\Delta\Phi(l)]\rangle &=& \biggl(\sum_{j=0}^{\infty}a_{2j}l^{2j}\biggr) + \biggl(\frac{N}{1+N\langle B_{0}^{2}\rangle / \langle B_{\rm{t}}^{2}\rangle}\biggr) \nonumber \\&&\times \biggl(\frac{1}{N_{1}}[1-e^{-l^{2}/2(\alpha^{2}+2W_{1}^{2})}]+ \frac{1}{N_{2}}[1-e^{-l^{2}/2(\alpha^{2}+2W_{2}^{2})}] - \frac{2}{N_{12}}[1-e^{-l^{2}/2(\alpha^{2}+W_{1}^{2}+W_{2}^{2})}]\biggr),
\end{eqnarray}
where $\Delta\Phi(l)$ is the angular difference of two line segments separated by a distance $l$, $\alpha$ is the turbulent correction length, the summation is the Taylor expansion of the ordered component of the ADF, $B_{\rm t}$ is the turbulent component of the {\em B}-field, $B_{0}$ is the ordered {\em B}-field, and $\delta\phi\sim\langle B_{\rm{t}}\rangle/\langle B_{\rm{0}}\rangle$.
$W_{1}$ and $W_{2}$ are the standard deviation (i.e., the FWHM divided by $\sqrt{8\ln2}$) of the synthesized beam and the LAS of the interferometric images, respectively, and $N$ is the number of turbulent cells along the line of sight probed by the telescope beam, expressed as:
\begin{eqnarray}\label{eq8}
N_{1}=\frac{(\alpha^{2}+2W_{1}^{2})\Delta^{\prime}}{\sqrt{2\pi}\alpha^{3}}, ~~~~
N_{2}=\frac{(\alpha^{2}+2W_{2}^{2})\Delta^{\prime}}{\sqrt{2\pi}\alpha^{3}}, ~~~~
N_{12}=\frac{(\alpha^{2}+W_{1}^{2}+W_{2}^{2})\Delta^{\prime}}{\sqrt{2\pi}\alpha^{3}}, ~~~~
N=\biggl(\frac{1}{N_{1}}+\frac{1}{N_{2}}-\frac{2}{N_{12}}\biggr)^{-1}.
\end{eqnarray}
Here $\Delta^{\prime}$ is the effective thickness of the region through which the signals are integrated along the line of sight.
It is estimated as the width at half of the maximum of the normalized autocorrelation function of the integrated normalized polarized flux \citep{houde2009dispersion}.
The values of $\Delta^{\prime}$ are listed in column 2 of Table \ref{Tab3:adf}.
The turbulent component of the ADF is expressed as:
\begin{eqnarray}\label{eq9}
b^{2}(l)=\biggl[\frac{N}{1+N\langle B_{0}^{2}\rangle / \langle B_{\rm{t}}^{2}\rangle}\biggr]\times \biggl[\frac{1}{N_{1}}e^{-l^{2}/2(\alpha^{2}+2W_{1}^{2})}+ \frac{1}{N_{2}}e^{-l^{2}/2(\alpha^{2}+2W_{2}^{2})} - \frac{2}{N_{12}}e^{-l^{2}/2(\alpha^{2}+W_{1}^{2}+W_{2}^{2})}\biggr].
\end{eqnarray}

\begin{table*}
\caption{Dispersion Analysis Results. Columns 2 to 8 present the effective thickness $\Delta^{\prime}$, turbulent correction length $\alpha$, square root of ratios of turbulent-to-ordered magnetic energy $(\langle B_{t}^{2}\rangle/\langle B_{0}^{2}\rangle)^{1/2}$, turbulent cells number $N$, correction factor $Q_{\rm u,~pos}^{\rm ADF}$, magnetic field strength derived from ADF method $B_{\rm u,~pos}^{\rm Hou16}$, and the number of Nyquist-Sampled {\em B}-field segments ($N_{\rm NS}$), respectively.}
\begin{center}
\label{Tab3:adf}
\begin{tabular}{l c c c c c c c c c c c c c c}
\hline
\hline
Name & ~ & $\Delta^{\prime}~(^{\prime\prime})$ & ~ & $\alpha~(^{\prime\prime})$ & ~ & $\biggl(\frac{\langle B_{t}^{2}\rangle}{\langle B_{0}^{2}\rangle}\biggr)^{\frac{1}{2}}$ & ~ & $N$ & ~ & $Q_{\rm u,~pos}^{\rm ADF}$ & ~ & $B_{\rm u,~pos}^{\rm Hou16}$ (mG) & ~ & $N_{\rm NS}$ \\
\hline
\textbf{Std-hourglass} \\
HOPS-87 & ~~~~~~ & 1.15 & ~~~~~~ & 1.03 & ~~~~~~ & 0.30 $\pm$ 0.03 & ~~~~~~ & 5.2 $\pm$ 1.4 & ~~~~~~ & 0.44 $\pm$ 0.08 & ~~~~~~ & 2.6 $\pm$ 1.1 & ~~~~~~ & 86 \\
HOPS-400 & ~~~ & 1.31 & ~ & 0.44 & ~ & 0.75 $\pm$ 0.08 & ~ & 3.2 $\pm$ 1.0 & ~ & 0.56 $\pm$ 0.06 & ~ & 1.0 $\pm$ 0.4 & ~ & 48 \\
HOPS-407 & ~~~ & 0.80 & ~ & 0.41 & ~ & 0.40 $\pm$ 0.04 & ~ & 2.5 $\pm$ 0.6 & ~ & 0.64 $\pm$ 0.05 & ~ & 1.7 $\pm$ 0.7 & ~ & 53 \\
\hline
\end{tabular}
\end{center}
\end{table*}

The lack of sufficient independent polarization measurements and the presence of random field orientations on scales of $\sim6^{\prime\prime}$ made the ADF method infeasible for most of the BOPS protostars. 
Additionally, \cite{hildebrand2009dispersion} and \cite{houde2009dispersion} suggest that at least 10--20 independent measurements are needed to reliably estimate the magnetic field properties.
We successfully applied this method only to three protostars.
Among these, HOPS-87 has sufficient statistics for the ADF fitting, while HOPS-400 and HOPS-407 lack sufficient independent measurements (see Figure \ref{fig:adf}).
The fitting results are listed in columns 3--5 of Table \ref{Tab3:adf}.
This method employs an analytical approach to determine the plane-of-sky polarization angles and derives the number of line-of-sight turbulent cells $N$, assuming that the line-of-sight turbulent correction scale is identical to the plane-of-sky turbulent correction scale.
Then, the amount of angular dispersion measured in polarization maps is reduced by a factor of $\sqrt{N}$ compared to the intrinsic angular dispersion \citep{hildebrand2009dispersion, cho2016dcf}, i.e., $Q_{\rm u,~pos}^{\rm ADF}=1/\sqrt{N}$ (see column 6 of Table \ref{Tab3:adf}).
Therefore, the plane-of-the-sky {\em B}-field strength, as listed in column 7 of Table \ref{Tab3:adf}, can be derived as \citep{houde2009dispersion}:
\begin{eqnarray}\label{eq10}
B_{\rm u,~pos}^{\rm Hou16}\simeq Q_{\rm u,~pos}^{\rm ADF}\sqrt{4\pi\rho}\sigma_{\rm nth}\biggl[\frac{\langle B_{t}^{2}\rangle}{\langle B_{\rm{0}}^{2}\rangle}\biggr]^{-1/2}.
\end{eqnarray}

\subsection{Mass-to-flux Ratio}

If the interstellar medium is well coupled to the {\em B}-field, the {\em B}-field can provide support against gravitational collapse.
The mass-to-flux ratio $\lambda$ characterizes the importance of magnetic fields relative to gravity and is described as \citep{crutcher2004scuba, liu2022magnetic}:
\begin{eqnarray}\label{eq11}
\lambda = \frac{(M/\Phi_{B})_{\rm obs}}{(M/\Phi_{B})_{\rm cr}} = \sqrt{4\pi^{2} G}\biggl[\frac{3}{2}\biggl(\frac{3-n}{5-2n}\biggr)\biggr]^{1/2} \frac{\mu_{\rm H_{2}}m_{\rm H}N_{\rm H_{2}}}{B} = 7.6\times10^{-21}\cdot\biggl[\frac{3}{2}\biggl(\frac{3-n}{5-2n}\biggr)\biggr]^{1/2}\cdot\frac{N_{\rm H_{2}}~[{\rm cm}^{-2}]}{B~[{\rm \mu G}^{-1}]]},
\end{eqnarray}
where $(M/\Phi_{B})_{\rm obs} = \mu_{\rm H_{2}}m_{\rm H}N_{\rm H_{2}}/B$ is the observed ratio of mass to flux, $(M/\Phi_{B})_{\rm cr} = \sqrt{k/6\pi^{2} G}$ is the critical value of mass to flux ratios and $k = (5-2n)/(3-n)$ is a correction factor for the density profile of $\rho(r)\propto r^{-n}$.
Column 8 of Table 2 lists the mass-to-flux ratios for targets with $\delta\phi\lesssim25^{\circ}$, calculated using the standard DCF method when $n = 2$, a typical index for the density profile of protostellar envelopes \citep[e.g.,][]{terebey1984collapse, andre2000PPIV}.
Columns 9--10 list the mass-to-flux ratios derived from the {\em B}-field strength calculated using the methods of Hei01 and Fal08, respectively.

\section{Discussion} \label{sec:discussions}

In Section \ref{sec:analysis}, we have derived the physical parameters and underlying {\em B}-field strengths for 26 BOPS protostellar envelopes with extended polarized emission within the central region around the protostar, with scales of inner $\sim$ 1200 au.
The uncertainties in the derived physical parameters arise from errors in factors such as dust opacities, dust temperatures, and observed fluxes.
As discussed in \cite{huang2024b}, we derived uncertainties of about 40\% for densities.
The uncertainty of the B-field strength listed in Table \ref{Tab2:strength} was derived from the uncertainties in polarization angle dispersion, non-thermal velocity dispersion, and density, using the error propagation method.
Indeed, \cite{liu2021dcf} found that the uncertainty of the {\em B}-field strength is a factor of $\sim2$ when the DCF assumption of energy balance holds.

In addition, the inclination angle of a disk relative to the line of sight influences the observed {\em B}-field morphology, primarily through projection effects, particularly in the context of an hourglass-shaped field structure around a protostar.
\cite{frau2011model} presented the magnetohydrodynamic collapse model and found that sources with hourglass (including std- and rot-hourglass) {\em B}-field structure are more easily observed when the disk inclination angle is greater than 30$^{\circ}$, especially in cases of edge-on disks ($\gtrsim60^{\circ}$).
In \cite{huang2024b}, we found that protostars with hourglass {\em B}-field structures can be observed with a disk inclination angle greater than 20$^{\circ}$.
Specifically, rot-hourglass protostars have larger disk inclination angles ($\gtrsim40^{\circ}$), while std-hourglass protostars have relatively small inclination angles ($20^{\circ}-40^{\circ}$).
This observed std-hourglass field with small inclination likely indicates that the {\em B}-field is sufficiently well-organized and strong.
However, protostars with std-hourglass {\em B}-field morphology often have small disk sizes close to the spatial resolution \citep{tobin2020vla, huang2024b}, thus higher angular resolution observations are required to better determine the disk inclination for compact disks.
In the following, we discuss the magnetized properties of the BOPS protostars.

\subsection{{The Distribution of \textit{B}-field Strength}}

The underlying {\em B}-field strengths are derived using the standard DCF method \citep{ostriker2001mag} for the 8 protostellar envelopes with {\em B}-field angle dispersion $\delta\phi$ less than 25$^{\circ}$.
In all these cases, the values derived are nearly identical to those using the Hei01 and Fal08 methods. 
Among these protostars, five exhibit a std-hourglass field shape, two exhibit a rot-hourglass morphology, and the remaining one shows the complex field configuration.
The field strengths for protostars with a std-hourglass field structure range from 0.6 to 2.9 mG, with a mean value of $\sim$ 2.0 mG (see column 5 of Table \ref{Tab2:strength}).
In contrast, the field strengths for the other three protostars appears to be smaller (0.5, 0.6 and 1.5 mG).
Regarding the ADF method, the {\em B}-field strength derived for one of the three protostars (HOPS-87) is similar to the value obtained using the standard DCF methods (see Tables \ref{Tab2:strength} and \ref{Tab3:adf}).
However, for the other two protostars (HOPS-400 and HOPS-407), the ADF method yields strengths of about half those returned by the other methods. 
This discrepancy may arise because the ADF fitting for these two cases was performed with statistics up to 4$^{\prime\prime}$, which is less than the scale of 6$^{\prime\prime}$ used to derive the physical parameters.
Additionally, the peak polarized emission of these BOPS protostars is very strong, and the effective thickness $\Delta^{\prime}$ may be underestimated, leading to lower values of $B_{\rm u,~pos}^{\rm Hou16}$.
The ADF method could potentially offer more accurate insights into the role of turbulence, but its limited applicability to BOPS protostars prevents the generalization of these results.

\begin{figure*}
\centering
\includegraphics[clip=true,trim=0.2cm 0.2cm 0.2cm 0.2cm,width=0.45 \textwidth]{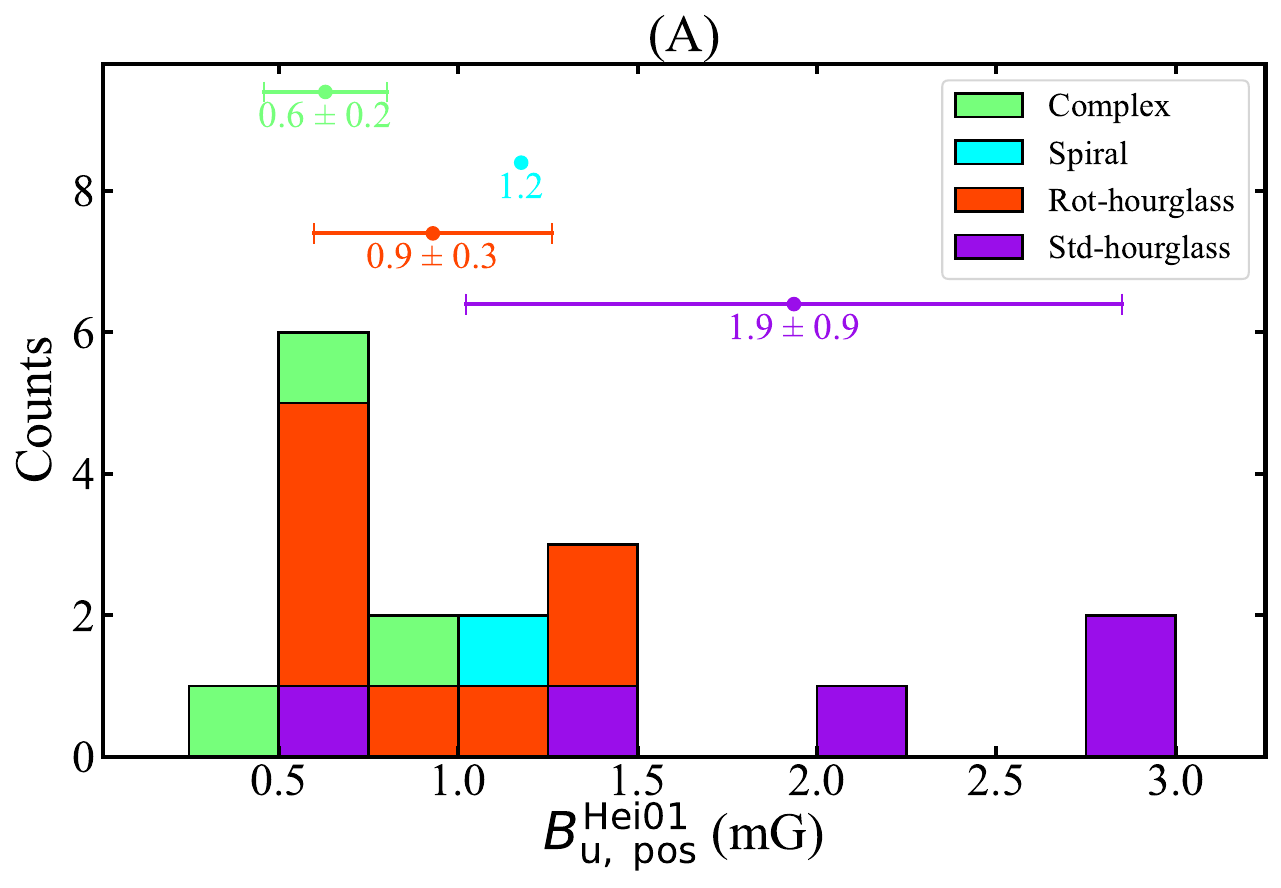}
~~~~~~~~~~
\includegraphics[clip=true,trim=0.2cm 0.2cm 0.2cm 0.2cm,width=0.45 \textwidth]{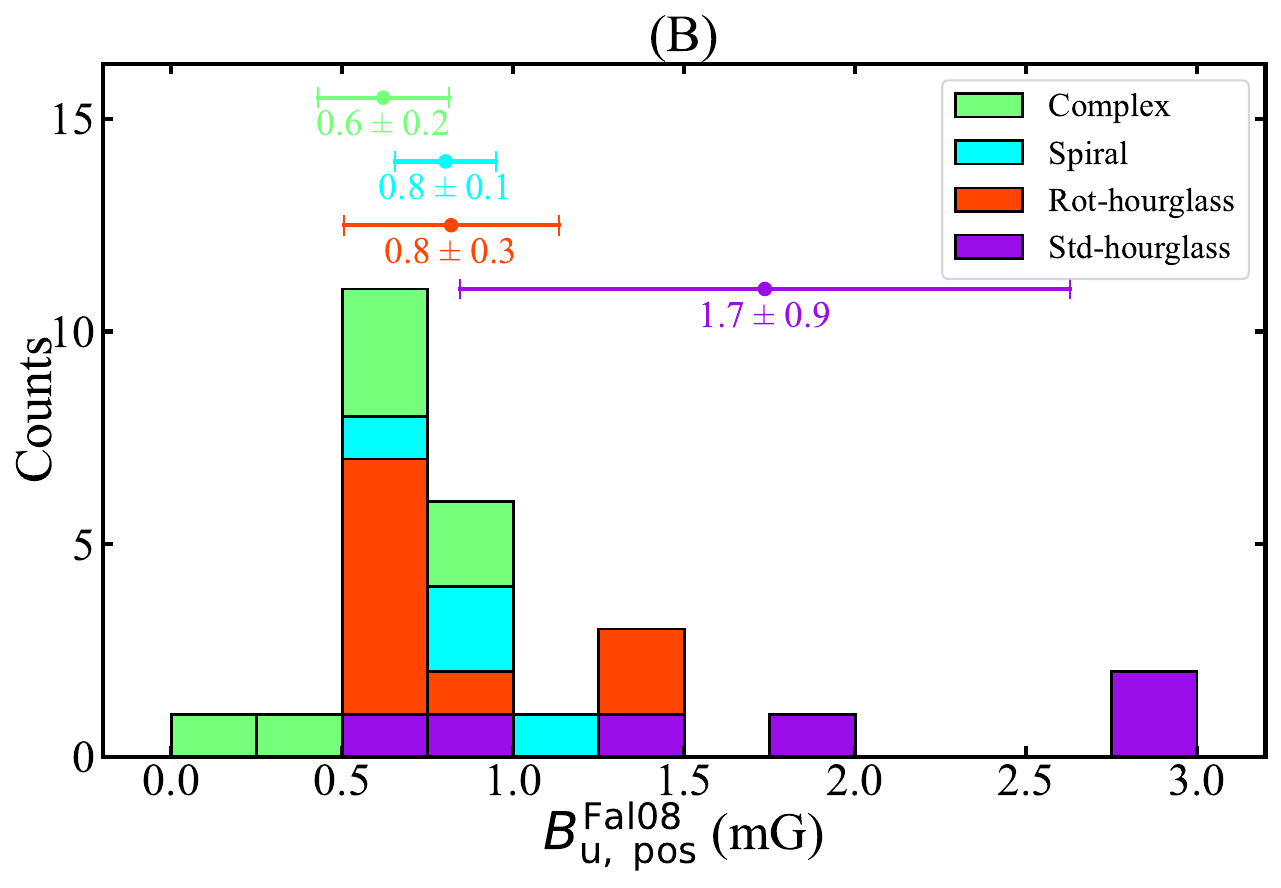}
~\\
~\\
\includegraphics[clip=true,trim=0.2cm 0.2cm 0.2cm 0.2cm,width=0.45 \textwidth]{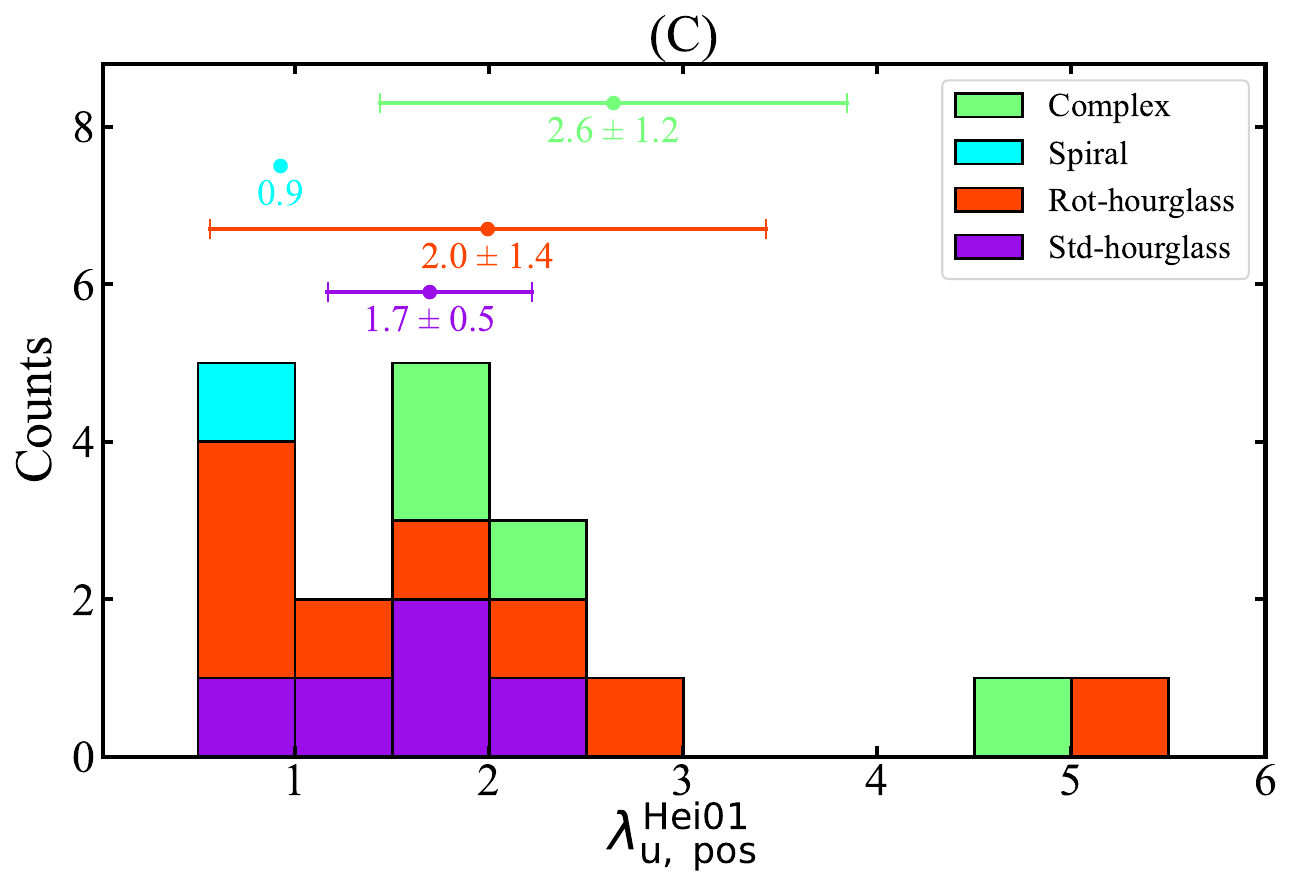}
~~~~~~~~~~
\includegraphics[clip=true,trim=0.2cm 0.2cm 0.2cm 0.2cm,width=0.45 \textwidth]{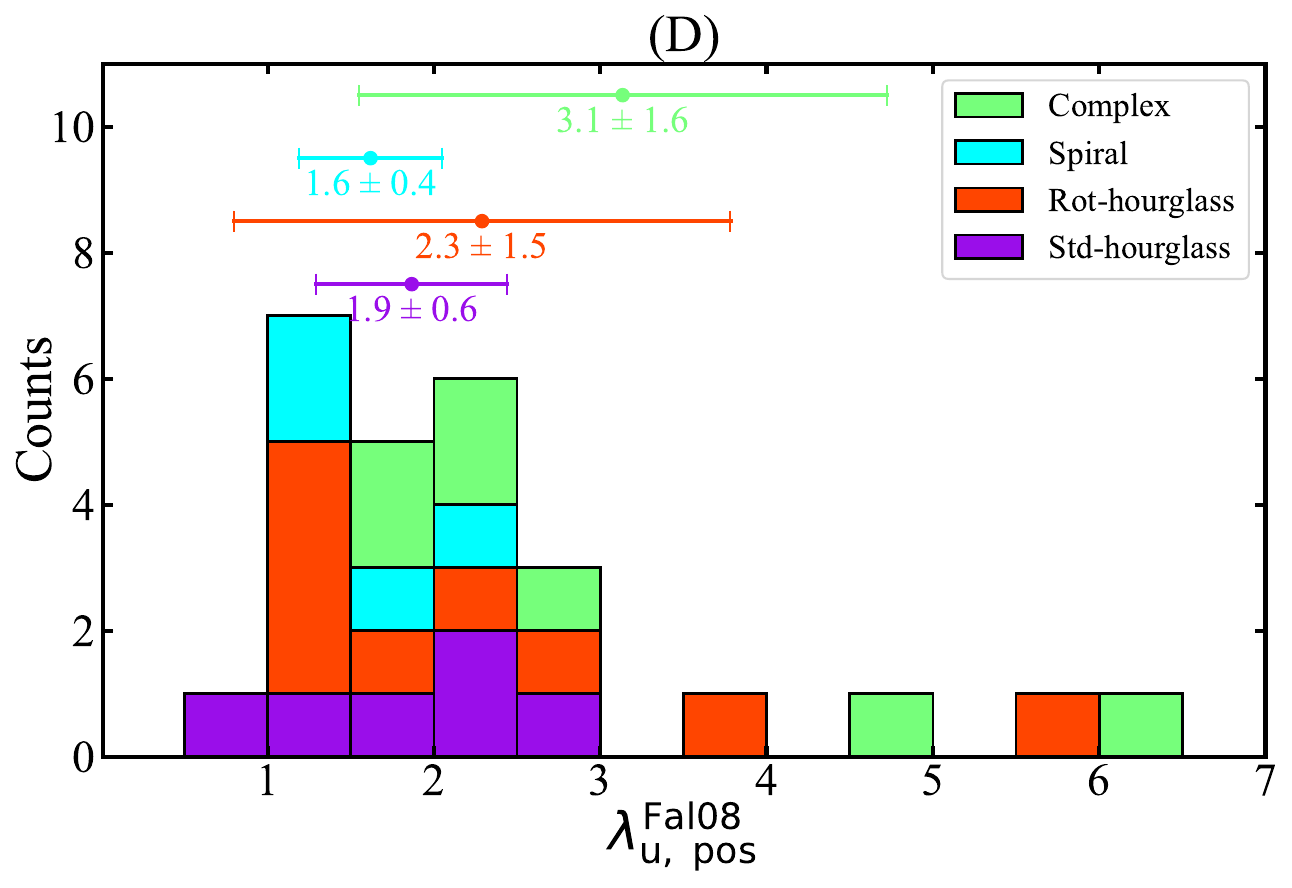}
\caption{The stacked histogram of $B_{\rm u,~pos}^{\rm Hei01}$ (Panel A) and $B_{\rm u,~pos}^{\rm Hei01}$ (Panels B), and of $\lambda_{\rm u,~pos}^{\rm Hei01}$ (panel C) and $\lambda_{\rm u,~pos}^{\rm Hei01}$ (panel D) for protostars with different {\em B}-field morphologies, respectively.
In panel A and panel C, the data for protostars with large uncertainties in $B_{\rm u,~pos}^{\rm Hei01}$ are not displayed here.
In all panels, the points indicate the mean values of the {\em B}-field strength, while the horizontal lines indicate the error bars derived from the standard deviation of statistics.
``Std-hourglass'', ``Rot-hourglass'', ``Spiral'', and ``Complex'' indicate protostars with different types of {\em B}-field structure of std-hourglass, rot-hourglass, spiral, and complex configurations, respectively.
} 
\label{fig:hist}
\end{figure*}

For protostars with $\delta\phi\gtrsim25^{\circ}$, $B_{\rm u,~pos}^{\rm Hei01}$ is generally similar to $B_{\rm u,~pos}^{\rm Fal08}$ in most cases.
However, in the Hei01 approach, when the term of ($\phi-\overline{\phi}$) is close to 90$^{\circ}$, the values derived from equation~\ref{eq5} is very small and associated with large uncertainties (larger than the values). 
This occurs for eight protostars (HOPS-12W, HOPS-169, HOPS-182, HOPS-359, HOPS-361N, HOPS-361S, HOPS-398, and OMC1N-6-7).
We excluded these protostars in the estimate of the mean magnetic field strength for each {\em B}-field class. 
Panels A and B of Figure~\ref{fig:hist} show the distributions of $B_{\rm u,~pos}^{\rm Hei01}$ and $B_{\rm u,~pos}^{\rm Fal08}$, respectively, with different colors representing the different types of {\em B}-fields.
We find that $B_{\rm u,~pos}^{\rm Hei01}$ for std-hourglass-field envelopes, with the mean value of $\sim1.9$ mG, is stronger than for envelopes with rot-hourglass, spiral, and complex {\em B}-field morphologies.
A similar trend is observed in panel B, where $B_{\rm u,~pos}^{\rm Fal08}$ for envelopes exhibiting std-hourglass, with a smaller mean field strength of $\sim1.7$ mG, is also stronger than for envelopes with other morphologies.
However, we find no significant differences between the {\em B}-field morphologies of rot-hourglass, spiral, and complex.
We note that the non-thermal velocity dispersions in spiral-{\em B}-field protostars are significantly larger than the mean value of $\sim 0.45~{\rm km~s^{-1}}$ \citep{huang2024b}. This may be due to the contribution of envelope/disk rotational motions to the line width. 
However, the limited spectral resolution does not allow us to properly test this hypothesis. 
In any case, this may lead to an overestimation of the {\em B}-field strength for spiral-{\em B}-field cases.

To assess the statistical reliability of our finding that std-hourglass protostars have higher magnetic field strengths compared to other morphologies, we performed two-sample Kolmogorov-Smirnov (K-S) tests for Hei01 and Fal08 methods, the p-values are listed in Table \ref{tab:pvalue}.
The K-S test results suggest that the difference in {\em B}-field strength between std-hourglass and other morphologies is statistically significant when considering the combined sample of non-std-hourglass morphologies, particularly in the Fal08 method ($P_{\rm Fal08}$ = 0.01).
However, when comparing std-hourglass to individual morphologies, the differences are not always statistically significant (e.g., rot-hourglass or spiral). This is likely due to the small sample sizes (especially for spiral) and the overlap in the distributions.
Additionally, when considering the uncertainties in the {\em B}-field strength estimates (denoted by $P_{\rm Hei01}^{\rm u}$ and $P_{\rm Fal08}^{\rm u}$), the p-values increase, underscoring the need for caution in interpreting these results.
Future studies with larger samples are necessary to confirm the observed trends and reduce the impact of uncertainties. 

\begin{table}
\centering
\caption{P-values from the K-S test comparing two distributions. The labels \textbf{std.}, \textbf{rot.}, \textbf{spi.}, and \textbf{com.} indicate magnetic field morphologies corresponding to the standard hourglass, rotated hourglass, spiral and complex structures, respectively. The label \textbf{oth.} encompasses all non-standard morphologies, including rotated hourglass, spiral, and complex structures.
The row labeled $P_{\rm Hei01}$ and $P_{\rm Fal08}$ show the p-values for the \cite{heitsch2001magnetic} and \cite{falceta2008DCF} methods, respectively, while the rows labeled $P_{\rm Hei01}^{\rm u}$ and $P_{\rm Fal08}^{\rm u}$ include the effects of uncertainties in the {\em B}-field strength estimates (as listed in Table \ref{Tab2:strength}).
The missing values in column 3 are due to the small sample size of spiral morphologies (only one case).}
\begin{tabular}{ccccccccc}
\hline
\hline
p-value & ~ & \textbf{std.} VS \textbf{rot.} & ~ & \textbf{std.} VS \textbf{spi.} & ~ & \textbf{std.} VS \textbf{com.} & ~ & \textbf{std.} VS \textbf{oth.} \\
\hline
$P_{\rm Hei01}$ & ~~~~~~~~~~~~~~~~~~~~~~~~ & 0.16 & ~~~~~~~~~~~~~~~~~~~~~~~~ & / & ~~~~~~~~~~~~~~~~~~~~~~~~ & 0.08 & ~~~~~~~~~~~~~~~~~~~~~~~~ & 0.05  \\
$P_{\rm Fal08}$ & ~ & 0.09 & ~ & 0.18 & ~ & 0.02 & ~ & 0.01 \\
\hline
$P_{\rm Hei01}^{\rm u}$ & ~ & 0.24 & ~ & / & ~ & 0.26 & ~ & 0.17 \\
$P_{\rm Fal08}^{\rm u}$ & ~ & 0.21 & ~ & 0.34 & ~ & 0.11 & ~ & 0.09 \\
\hline
\end{tabular}
\label{tab:pvalue}
\end{table}

The differences in {\em B}-field strength between std-hourglass and other morphologies are primarily driven by variations in $\delta\phi$ rather than density.
Specifically, protostars with std-hourglass morphologies tend to exhibit smaller $\delta\phi$ values (with mean value of 18.5$^{\circ}$), which is approximately half the values observed for rotated hourglass, spiral, or complex configurations (with mean value of 34.1$^{\circ}$). 
This is consistent with the idea that the std-hourglass morphology reflects a stronger, more coherent {\em B}-field that is better able to resist turbulent perturbations. 
In contrast, the gas densities of std-hourglass protostars, while generally higher than those of other morphologies, do not show a sufficiently large spread to fully account for the observed differences in {\em B}-field strength. 
For example, the mean density of std-hourglass protostars is approximately
$6.1\times10^{-17}~{\rm g~cm}^{-3}$, compared to $3.8\times10^{-17}~{\rm g~cm}^{-3}$ for other morphologies. 
The factor for $\sqrt{\rho}$ between std-hourglass cases and other cases is $\sim1.2$. 
This modest difference in density cannot alone explain the factor of $\sim 2$ difference in magnetic field strength between std-hourglass and other configurations.
Therefore, the polarization measurements of $\delta\phi$, rather than variations in density, are the primary driver of the higher magnetic field strengths observed in std-hourglass protostars.

\cite{huang2024magnetic} found that protostars exhibiting std-hourglass morphologies are associated with unresolved velocity gradients ($\lesssim1.0~{\rm km~s^{-1}~arcsec^{-1}}$), whereas approximately half of the protostellar envelopes with rot-hourglass and over 60\% of those with spiral-{\em B}-field configurations exhibit significant velocity gradients ($\gtrsim1.0~{\rm km~s^{-1}~arcsec^{-1}}$).
These findings suggest that the {\em B}-field in std-hourglass configuration appears to be strong enough to effectively slow down the rotational rate, while the {\em B}-field strength in rot-hourglass and spiral configurations is likely to be less significant.
Our results show that {\em B}-field strengths associated with std-hourglass morphologies are stronger than those in other configurations, consistent with previous studies.

\subsection{The Distribution of Mass-to-flux Ratio}

Figure \ref{fig:hist} shows the distribution and the mean values of $\lambda_{\rm u,~pos}^{\rm Hei01}$ (panel C) and $\lambda_{\rm u,~pos}^{\rm Fal08}$ (panel D) for the different {\em B}-field morphologies. 
For $\lambda_{\rm u,~pos}^{\rm Hei01}$, we have excluded 8 protostars with large uncertainties in the magnetic field strength (see Section~2.2 and 3.1). 
Contrary to the trend observed for {\em B}-field strength, we do not find significant differences in the mean mass-to-flux ratio for the different {\em B}-field morphologies. 
Most protostars have mass-to-flux ratios between 1 and 3, i.e. slightly supercritical. 
However, we note that all protostellar envelopes exhibiting a std-hourglass {\em B}-field have mass-to-flux ratio below 3, while 4 protostars with other morphologies with mass-to-flux ratio larger than 3. 
In MHD simulations, the mass-to-flux ratio represents the initial value before any protostar has formed. 
However, most observational studies are focused on star-forming regions, where the mass already accreted onto the protostars should also be included. 
For the BOPS sample, there are only two reported estimates of the protostellar mass: 2.5 and 0.27~M$_{\odot}$ for HOPS-370 and HH212M, respectively \citep{lee2017hh212, tobin2020b}. These are 12 and 0.9 times the envelope masses within 1200~au (0.21 and 0.31~M$_{\odot}$), respectively. 
In the case of HOPS-370, this means that the true mass-to-flux ratio is $\sim 12$, an order of magnitude larger than the value reported here with only the envelope mass (HH212M is not part of the sample in this paper). 
Therefore, the mass-to-flux ratio derived using only the envelope mass should be used with caution, since they may be significantly underestimated. 

Recently, \cite{nacho2024magnetic} used non-ideal MHD models of protostellar disk formation and evolution to search for evidence of magnetic braking.
They suggest that strongly and weakly magnetized models tend to exhibit a std-hourglass and rot-hourglass {\em B}-field morphology, respectively. 
They considered strongly- and weakly-magnetized models as those with $\lambda\lesssim3$ and $\lambda\gtrsim3$, respectively. 
Despite the uncertainties in the derived $\lambda$, our results appear to be consistent with the simulation predictions in the case of a std-hourglass field morphology.
Furthermore, a slight trend from some cases (HOPS-78, HOPS-169 and HOPS-317S) is observed that rot-hourglass field configurations would poss larger mass-to-flux ratios and would be less magnetized compared to std-hourglass field morphologies.

\section{Conclusions} \label{sec:con}

The BOPS presented the largest polarization study of magnetic fields in 61 young protostellars, and classified them into four major types of magnetic field morphology: standard hourglass, rotated hourglass, spiral, and complex configuration.
Among the BOPS sample, 26 protostars exhibit extended polarized emission at least $6^{\prime\prime}$, corresponding to scales of $\sim$ 2400 au.
In this study, we focus on these 26 protostars within the inner $\sim$ 2400 au scales \citep[see][]{huang2024magnetic, huang2024b}.
We used the standard DCF method \citep{ostriker2001mag} to estimate the magnetic field strength for 8 protostars with polarization angle dispersions $\lesssim25^{\circ}$. 
Additionally, we applied the DCF method with corrections proposed by Hei01 and Fal08 for 26 protostars.
The ADF method was successfully applied to 3 protostars with standard hourglass magnetic field structures.
The main results are as follows:

\begin{enumerate}
\item For protostars with standard hourglass magnetic field morphology, the magnetic field strengths derived from the DCF method with different corrections are similar (differing by less than 0.1 mG), with an average value of $\sim 2.0\pm0.9$ mG.

\item For protostars with rotated hourglass, spiral, and complex magnetic field morphologies, the magnetic field strengths ($\lesssim1.0$~mG) are generally lower than those in the case of standard hourglass, suggesting that the magnetic field is stronger in protostellar envelopes that exhibit a standard hourglass field morphology.

\item The field strength derived from the ADF method for HOPS-87 is consistent with those obtained using the other DCF methods. However, for HOPS-400 and HOPS-407, the ADF method yields strengths that are approximately half those derived from the other methods, likely due to an underestimation of the effective thickness and insufficient statistics for ADF fitting.

\item Most protostars are slightly supercritical, with mass-to-flux ratios between 1.0 and 3.0.
In particular, all protostars with a standard hourglass magnetic field morphology have mass-to-flux ratios $<3.0$, whereas 4 out of 20 with other field morphologies have values higher than 3.0.
However, these mass-to-flux ratios do not include contributions from protostellar mass, so they are likely significantly underestimated, at least in some cases.
\end{enumerate}

\begin{acknowledgments}
B.H., J.M.G. and A.S.-M. acknowledge support by the grant PID2020-117710GB-I00 and PID2023-146675NB-I00 (MCI-AEI-FEDER, UE). 
B.H. acknowledges financial support from the China Scholarship Council (CSC) under grant No. 202006660008.
This work is also partially supported by the program Unidad de Excelencia María de Maeztu CEX2020-001058-M.
A.S.-M. acknowledges support from the RyC2021-032892-I grant funded by MCIN/AEI/10.13039/501100011033 and by the European Union `Next GenerationEU’/PRTR.
L.W.L acknowledges support from NSF AST-1910364 and NSF AST-2307844.
M.F.L. thanks the hospitality and the financial support of the Instituto de Radioastronom\'ia y Astrof\'isica (UNAM, Morelia, M\'exico).
W.K. was supported by the National Research Foundation of Korea (NRF) grant funded by the Korea government (MSIT) (RS-2024-00342488).
Z.Y.L. is supported in part by NASA 80NSSC20K0533 and NSF AST-2307199.
The authors acknowledge Anaëlle Maury for helpful discussions and Jacob Labonte for early analysis of the BOPS data.
This paper makes use of the following ALMA data: ADS/JAO.ALMA\#2019.1.00086.
ALMA is a partnership of ESO (representing its member states), NSF (USA) and NINS (Japan), together with NRC (Canada), MOST and ASIAA (Taiwan), and KASI (Republic of Korea), in cooperation with the Republic of Chile. The Joint ALMA Observatory is operated by ESO, AUI/NRAO and NAOJ.
\end{acknowledgments}


\begin{thebibliography}{}
\bibitem[{A{\~n}ez-L{\'o}pez {et~al.}(2024)A{\~n}ez-L{\'o}pez, {et~al.}}]{nacho2024magnetic}
A{\~n}ez-L{\'o}pez, N., Lebreuilly, U., Maury, A., {et~al.} 2024, \aap, 687, A63

\bibitem[{Andersson {et~al.}(2015)Andersson, {et~al.}}]{andersson2015interstellar}
Andersson, B. G., Lazarian, A., \& Vaillancourt, J. E., 2015, \araa, 53, 501

\bibitem[{Andre {et~al.}(2000)Andre, {et~al.}}]{andre2000PPIV}
Andre, P., Ward-Thompson, D., \& Barsony, M., 2000, in Protostars and Planets IV, ed. V. Mannings, A. P. Boss, \& S. S. Russell, 59

\bibitem[{Chandrasekhar {et~al.}(1953)Chandrasekhar, {et~al.}}]{chandra1953magnetic}
Chandrasekhar, S., \& Fermi, E., 1953, \apj, 118, 116

\bibitem[{Chen {et~al.}(2022)Chen, {et~al.}}]{chen2022dcf}
Chen, C. -Y., Li, Z. -Y., Mazzei, R. R., {et~al.} 2022, \mnras, 514, 1575

\bibitem[{Cho {et~al.}(2016)Cho, {et~al.}}]{cho2016dcf}
Cho, J., \& Yoo, H. 2016, \apj, 821, 21

\bibitem[{Cortes {et~al.}(2021)Cortes, {et~al.}}]{cortes2021magnetic}
Cort{\'e}s, P. C., Sanhueza, P., Houde, M., {et~al.} 2021, \apj, 923, 204

\bibitem[{Cox {et~al.}(2018)Cox, {et~al.}}]{cox2018alma}
Cox, E. G., Harris, R. J., Looney, L. W., {et~al.} 2018, \apj, 855, 92

\bibitem[{Crutcher(1999)Crutcher}]{crutcher1999magnetic}
Crutcher, R. M. 1999, \apj, 520, 706

\bibitem[{Crutcher {et~al.}(2009)Crutcher, {et~al.}}]{crutcher2009magnetic}
Crutcher, R. M., Hakobian, N., \& Troland, T. H. 2009, \apj, 692, 844

\bibitem[{Crutcher {et~al.}(2004)Crutcher, {et~al.}}]{crutcher2004scuba}
Crutcher, R. M., Nutter, D. J., Ward-Thompson, D., {et~al.} 2004, \apj, 600, 279

\bibitem[{Davis(1951)Davis}]{davis1951dcf}
Davis, L. 1951, Phys. Rev., 81, 890

\bibitem[{Falceta-Gon{\c{c}}alves {et~al.}(2016)Falceta-Gon{\c{c}}alves, {et~al.}}]{falceta2008DCF}
Falceta-Gon{\c{c}}alves, D., Lazarian, A., \& Kowal, G. 2008, \apj, 679, 537

\bibitem[{Frau {et~al.}(2011)Frau, {et~al.}}]{frau2011model}
Frau, P., Galli, D., \& Girart, J. M. 2011, \aap, 535, A44

\bibitem[{Galametz {et~al.}(2018)Galametz, {et~al.}}]{galametz2018sma}
Galametz, M., Maury, A., Girart, J. M., {et~al.} 2018, \aap, 616, A139

\bibitem[{Girart {et~al.}(2009)Girart, {et~al.}}]{girart2009magnetic}
Girart, J. M., Beltr{\'a}n, M. T., Zhang, Q., {et~al.} 2009, Sci, 324, 1408

\bibitem[{Girart {et~al.}(1999)Girart, {et~al.}}]{girart1999detection}
Girart, J. M., Crutcher, R. M., \& Rao, R. 1999, \apjl, 525, L109

\bibitem[{Girart {et~al.}(2013)Girart, {et~al.}}]{girart2013dr}
Girart, J. M., Frau, P., Zhang, Q., {et~al.} 2013, \apj, 772, 69

\bibitem[{Girart {et~al.}(2006)Girart, {et~al.}}]{girart2006magnetic}
Girart, J. M., Rao, R., \& Marrone, D. P. 2006, Sci, 313, 812

\bibitem[{Heitsch {et~al.}(2001)Heitsch, {et~al.}}]{heitsch2001magnetic}
Heitsch, F., Zweibel, E. G., Mac Low, M.-M., {et~al.} 2001, \apj, 561, 800

\bibitem[{Hildebrand {et~al.}(2009)Hildebrand, {et~al.}}]{hildebrand2009dispersion}
Hildebrand, R. H., Kirby, L., Dotson, J. L., {et~al.} 2009, \apj, 696, 567

\bibitem[{Hoang {et~al.}(2009)Hoang, {et~al.}}]{hoang2009grain}
Hoang, T., \& Lazarian, A. 2009, \apj, 697, 1316

\bibitem[{Houde {et~al.}(2016)Houde, {et~al.}}]{houde2016dispersion}
Houde, M., Hull, C. L. H., Plambeck, R. L., {et~al.} 2016, \apj, 820, 38

\bibitem[{Houde {et~al.}(2009)Houde, {et~al.}}]{houde2009dispersion}
Houde, M., Vaillancourt, J. E., Hildebrand, R. H., {et~al.} 2009, \apj, 706, 1504

\bibitem[{Huang {et~al.}(2024)Huang, {et~al.}}]{huang2024magnetic}
Huang, B., Girart, J. M., Stephens, I. W., {et~al.} 2024, \apjl, 963, L31

\bibitem[{Huang {et~al.}(2025)Huang, {et~al.}}]{huang2024b}
Huang, B., Girart, J. M., Stephens, I. W., {et~al.} 2025, \apj, 981, 30

\bibitem[{Hull {et~al.}(2020)Hull, {et~al.}}]{hull2020understanding}
Hull, C. L. H., Le Gouellec, V. J. M., Girart, J. M., {et~al.} 2020, \apj, 892, 152

\bibitem[{Hull {et~al.}(2019)Hull, {et~al.}}]{hull2019interferometric}
Hull, C. L. H., \& Zhang, Q. 2019, Frontiers in Astronomy and Space Sciences, 6, 3

\bibitem[{Inoue {et~al.}(2013)Inoue, {et~al.}}]{inoue2013ins}
Inoue, T., \& Fukui, Y. 2013, \apjl, 774, L31

\bibitem[{Kauffmann {et~al.}(2008)Kauffmann, {et~al.}}]{kauffmann2008mambo}
Kauffmann, J., Bertoldi, F., Bourke, T. L., {et~al.} 2008, \aap, 487, 993

\bibitem[{Kounkel {et~al.}(2017)Kounkel, {et~al.}}]{Kounkel2017dis}
Kounkel, M., Hartmann, L., Loinard, L., {et~al.} 2017, \apj, 834, 142

\bibitem[{Kwon {et~al.}(2022)Kwon, {et~al.}}]{kwon2022obs}
Kwon, W., Pattle, K., Sadavoy, S., {et~al.} 2022, \apj, 926, 163

\bibitem[{Le Gouellec {et~al.}(2019)Le Gouellec, {et~al.}}]{le2019characterizing}
Le Gouellec, V. J. M., Hull, C. L. H., Maury, A. J., {et~al.} 2019, \apj, 885, 106

\bibitem[{Le Gouellec {et~al.}(2020)Le Gouellec, {et~al.}}]{le2020IMS}
Le Gouellec, V. J. M., Maury, A. J., Guillet, V. {et~al.} 2020, \aap, 644, A11

\bibitem[{Lee {et~al.}(2017)Lee, {et~al.}}]{lee2017hh212}
Lee, C.-F., Li, Z.-Y., Ho, P. T. P. {et~al.} 2017, \apj, 843, 27

\bibitem[{Li {et~al.}(2022)Li, {et~al.}}]{li2022magnetic}
Li, P. S., Lopez-Rodriguez, E., Ajeddig, H. {et~al.} 2022, \mnras, 510, 6085

\bibitem[{Liu {et~al.}(2022a)Liu, {et~al.}}]{liu2022magnetic}
Liu, J., Qiu, K., \& Zhang, Q. 2022a, \apj, 925, 30

\bibitem[{Liu {et~al.}(2021)Liu, {et~al.}}]{liu2021dcf}
Liu, J., Zhang, Q., Commer{\c c}on, B., {et~al.} 2021, \apj, 919, 79

\bibitem[{Liu {et~al.}(2022b)Liu, {et~al.}}]{liu2022dcf}
Liu, J., Zhang, Q., \& Qiu, K. 2022b, Frontiers in Astronomy and Space Sciences, 9, 943556

\bibitem[{Mahieu {et~al.}(2012)Mahieu, {et~al.}}]{Mahieu2012alma}
Mahieu, S., Maier, D., Lazareff, B., {et~al.} 2012, IEEE Transactions on Terahertz Science and Technology, 2, 29

\bibitem[{Maury {et~al.}(2022)Maury, {et~al.}}]{maury2022}
Maury, A., Hennebelle, P., \& Girart, J. M. 2022, Frontiers in Astronomy and Space Sciences, 9, 949223

\bibitem[{Moechel {et~al.}(2015)Moechel, {et~al.}}]{moechel2015ism}
Moechel, N., \& Burkert, A. 2015, \apj, 807, 67

\bibitem[{Myers {et~al.}(2024)Myers, {et~al.}}]{myers2024DCF}
Myers, P. C., Stephens, I. W., \& Coud{\'e}, S. 2024, \apj, 962, 64

\bibitem[{Ostriker {et~al.}(2001)Ostriker, {et~al.}}]{ostriker2001mag}
Ostriker, E. C., Stone, J. M., \& Gammie, C. F. 2001, \apj, 546, 980

\bibitem[{Padoan {et~al.}(2001)Padoan, {et~al.}}]{padoan2001ism}
Padoan, P., Juvela, M., Goodman, A. A., {et~al.} 2001, \apj, 553, 227

\bibitem[{Padoan {et~al.}(1999)Padoan, {et~al.}}]{padoan1999cloud}
Padoan, P., \& Nordlund, Å. 1999, \apj, 526, 279

\bibitem[{Pattle {et~al.}(2023)Pattle, {et~al.}}]{pattle2023ppvii}
Pattle, K., Fissel, L., Tahani, M., {et~al.} 2023, in Astronomical Society of the Pacific Conference Series, Vol. 534, Protostars and Planets VII, ed. S. Inutsuka, Y. Aikawa, T. Muto, K. Tomida, \& M. Tamura, 193

\bibitem[{Qiu {et~al.}(2014)Qiu, {et~al.}}]{qiu2014sma}
Qiu, K., Zhang, Q., Menten, K. M., {et~al.} 2014, \apjl, 794, L18

\bibitem[{Ram{\'i}rez-Galeano {et~al.}(2022)Ram{\'i}rez-Galeano, {et~al.}}]{ramirez2022gravity}
Ram{\'i}rez-Galeano, L., Ballesteros-Paredes, J., Smith, R. J., {et~al.} 2022, \mnras, 515, 2822

\bibitem[{Skalidis {et~al.}(2021)Skalidis, {et~al.}}]{skalidis2021dcf}
Skalidis, R., \& Tassis, K. 2021, \aap, 647, A186

\bibitem[{Stephens {et~al.}(2013)Stephens, {et~al.}}]{stephens2013hourglass}
Stephens, I. W., Looney, L. W., Kwon, W., {et~al.} 2013, \apjl, 769, L15

\bibitem[{Terebey {et~al.}(1984)Terebey, {et~al.}}]{terebey1984collapse}
Terebey, S., Shu, F. H., \& Cassen, P. 1984, \apj, 286, 529

\bibitem[{Tobin {et~al.}(2020a)Tobin, {et~al.}}]{tobin2020vla}
Tobin, J. J., Sheehan, P. D., Megeath, S. T., {et~al.} 2020a, \apj, 890, 130

\bibitem[{Tobin {et~al.}(2020b)Tobin, {et~al.}}]{tobin2020b}
Tobin, J. J., Sheehan, P. D., Reynolds, N., {et~al.} 2020b, \apj, 905, 162

\bibitem[{Van Loo {et~al.}(2014)Van Loo, {et~al.}}]{van2014ism}
Van Loo, S., Keto, E., \& Zhang, Q. 2014, \apj, 789, 37

\bibitem[{V{\'a}zquez-Semadeni {et~al.}(2019)V{\'a}zquez-Semadeni, {et~al.}}]{vazquez2019ghc}
V{\'a}zquez-Semadeni, E., Palau, A., Ballesteros-Paredes, J., {et~al.} 2019, \mnras, 490, 3061

\bibitem[{Zhang {et~al.}(2014)Zhang, {et~al.}}]{zhang2014sma}
Zhang, Q., Qiu, K., Girart, J. M., {et~al.} 2014, \apj, 792, 116

\end{thebibliography}
\end{document}